\documentclass[12pt]{article}
\usepackage{amsmath,amssymb}
\newcommand{\eqdef}{\stackrel{\text{def}}{=}}
\newcommand{\n}{\nonumber\\}
\newcommand{\bm}{\boldsymbol}
\newcommand{\ignore}[1]{}

\newcommand{\Romannumeral}[1]{\uppercase\expandafter{\romannumeral#1}}
\newcommand{\I}{\text{\Romannumeral{1}}}
\newcommand{\II}{\text{\Romannumeral{2}}}

\allowdisplaybreaks[4]

\begin{document}

\baselineskip=20pt

%%%%%%%%%%%%%%%%%%%%%%%%%%%%%%%%%%%%%%%%%%%%%%%%%%%%%%%%%%%%
%                                                          %
%  Title page                                              %
%                                                          %
%%%%%%%%%%%%%%%%%%%%%%%%%%%%%%%%%%%%%%%%%%%%%%%%%%%%%%%%%%%%
\newfont{\elevenmib}{cmmib10 scaled\magstep1}
\newcommand{\preprint}{
    \begin{flushleft}
     \elevenmib Yukawa\, Institute\, Kyoto\\
   \end{flushleft}\vspace{-1.3cm}
   \begin{flushright}\normalsize \sf
     DPSU-12-1\\
     YITP-12-18\\
%     {\tt arXiv:1203.5868[math-ph]}\\
%     March 2012
   \end{flushright}}
\newcommand{\Title}[1]{{\baselineskip=26pt
   \begin{center} \Large \bf #1 \\ \ \\ \end{center}}}
\newcommand{\Author}{\begin{center}
   \large \bf Satoru Odake${}^a$ and Ryu Sasaki${}^b$ \end{center}}
\newcommand{\Address}{\begin{center}
     $^a$ Department of Physics, Shinshu University,\\
     Matsumoto 390-8621, Japan\\
     ${}^b$ Yukawa Institute for Theoretical Physics,\\
     Kyoto University, Kyoto 606-8502, Japan
   \end{center}}
\newcommand{\Accepted}[1]{\begin{center}
   {\large \sf #1}\\ \vspace{1mm}{\small \sf Accepted for Publication}
   \end{center}}

\preprint
\thispagestyle{empty}

\Title{Multi-indexed ($q$-)Racah Polynomials}

\Author

\Address
\vspace{1cm}

\begin{abstract}
As the second stage of the project {\em multi-indexed orthogonal
polynomials\/}, we present, in the framework of `discrete quantum
mechanics' with real shifts in one dimension, the multi-indexed
($q$-)Racah polynomials. They are obtained from the ($q$-)Racah
polynomials by multiple application of the discrete analogue of
the Darboux transformations or the Crum-Krein-Adler deletion of
`virtual state' vectors, in a similar way
to the multi-indexed Laguerre and Jacobi polynomials reported earlier.
The virtual state vectors are the `solutions' of the matrix
Schr\"odinger equation with negative `eigenvalues', except for
one of the two boundary points.
\end{abstract}

%%%%%%%%%%%%%%%%%%%%%%%%%%%%%%%%%%%%%%%%%%%%%%%%%%%%%%%%%%%%%%%
%                                                             %
%  1. Introduction                                            %
%                                                             %
%%%%%%%%%%%%%%%%%%%%%%%%%%%%%%%%%%%%%%%%%%%%%%%%%%%%%%%%%%%%%%%
\section{Introduction}
\label{sec:intro}

This is a second report of the project {\em multi-indexed orthogonal
polynomials\/}. Following the examples of multi-indexed Laguerre and
Jacobi polynomials \cite{os25}, multi-indexed ($q$-)Racah polynomials
are constructed in the framework of discrete quantum mechanics with real
shifts \cite{os12}. It should be emphasised that the original ($q$-)Racah
polynomials are the most generic members of the Askey scheme of
hypergeometric orthogonal polynomials with purely discrete orthogonality
measures \cite{askey, ismail,koeswart,gasper}. They are also called
orthogonal polynomials of a discrete variable \cite{nikiforov}.
These new multi-indexed orthogonal polynomials are specified by a set of
indices $\mathcal{D}=\{d_1,\ldots,d_M\}$ consisting of distinct natural
numbers $d_j\in\mathbb{N}$, on top of $n$, which counts the nodes as in
the ordinary orthogonal polynomials. The simplest examples,
$\mathcal{D}=\{\ell\}$, $\ell\ge1$, $\{P_{\ell,n}(x)\}$ are also called
{\em exceptional orthogonal polynomials\/} \cite{gomez}--\cite{quesne3}.
They are obtained as the main part of the eigenfunctions (vectors) of
various {\em exactly solvable\/} Schr\"odinger equations in one dimensional
quantum mechanics and their `discrete' generalisations, in which the
corresponding Schr\"odinger equations are second order difference equations
\cite{os12,os13,os14}. They form a complete set of orthogonal polynomials,
although they start at a certain positive degree ($\ell\ge1$) rather than
a degree zero constant term. The latter situation is essential for avoiding
the constraints of Bochner's theorem \cite{bochner}. The exceptional
Laguerre polynomials with two extra indices $\mathcal{D}=\{d_1,d_2\}$ were
introduced in \cite{gomez3}. We are quite sure that these new orthogonal
polynomials will find plenty of novel applications in various branches
of science and technology as other orthogonal polynomials.
One obvious application is the birth and death processes \cite{bdp}.
These new orthogonal polynomials provide huge stocks of {\em exactly
solvable birth and death processes\/} \cite{bdproc}. The transition
probabilities are given explicitly, not in a general spectral
representation form of Karlin-McGregor \cite{karlin}. An interesting
possible application is to one-dimensional spin systems and 
quantum information theory \cite{vinet}.

The basic logic for constructing multi-indexed orthogonal polynomials
is essentially the same for the ordinary Schr\"odinger equations,
{\em i.e.\/} those for the Laguerre and Jacobi polynomials and for
the difference Schr\"odinger equations with real shifts, {\em i.e.\/}
the ($q$-)Racah polynomials, etc. The main ingredients are the factorised
Hamiltonians, the Crum-Krein-Adler formulas \cite{crum,adler,Nsusy} for
deletion of eigenstates, {\em that is\/} the multiple Darboux
transformations \cite{darb} and the virtual states solutions \cite{os25}
which are generated by twisting the discrete symmetries of the original
Hamiltonians. Most of these methods for discrete Schr\"odinger equations
had been developed \cite{os12,os24,os13,os14,os15,gos,os22} and they
were used for the exceptional ($q$-)Racah polynomials \cite{os23}.
The concept of virtual state `solutions' requires special explanation
in the present case. In the ordinary quantum mechanics cases, the virtual
state solutions are the solutions of the Schr\"odinger equation but
they do not belong to the Hilbert space of square integrable functions
due to the twisted boundary conditions. In the present case, the
Hamiltonians are finite-dimensional real symmetric tri-diagonal matrices.
Therefore the eigenvalue equations for a given Hamiltonian matrix cannot
have any extra solutions other than the genuine eigenvectors. Thus we
will use the term {\em virtual state vectors\/}. As will be shown in the
text, virtual state vectors are the `solutions' of the eigenvalue problem
for a {\em virtual\/} Hamiltonian $\mathcal{H}'$, except for one of the
boundaries, $x=x_{\text{max}}$ \eqref{schreq'}.
The virtual Hamiltonians are obtained from the original Hamiltonian by
twisting the discrete symmetry and they are linearly related to the
original Hamiltonian \eqref{HH'}.
Thus the virtual state vectors `satisfy' the eigenvalue equation for
the original Hamiltonian, except for one of the two boundaries.
The polynomial part of the virtual state vectors had been used
for the exceptional ($q$-)Racah polynomials.
One distinctive feature of virtual states deletion in discrete quantum
mechanics with real shifts is that the size of the Hamiltonian matrix
($x_{\text{max}}$) remains the same. This is in marked contrast with
the eigenstates deletion (Christoffel transformations \cite{askey,os22}),
in which case the size decreases by the number of deleted eigenstates.

This paper is organised as follows.
In section two, the main ingredients of the theory, the difference
Schr\"odinger equation for the ($q$-)Racah system, the polynomial
eigenvectors and virtual state vectors, are introduced.
Starting from the general setting of discrete quantum mechanics with
real shifts in \S\,\ref{sec:form}, the basic properties of the ($q$-)Racah
systems are recapitulated in \S\,\ref{sec:org_qR}. Based on the twisting
(symmetry), the virtual state vectors are introduced in \S\,\ref{sec:virtual}.
Section three is the main part of the paper.
The basic logic of virtual states deletion in discrete quantum mechanics
with real shifts in general is outlined in \S\,\ref{sec:virtual_del}.
The explicit forms of multi-indexed ($q$-)Racah polynomials are provided
in \S\,\ref{sec:ef_miop_qR}. The final section is for a summary and comments.

%%%%%%%%%%%%%%%%%%%%%%%%%%%%%%%%%%%%%%%%%%%%%
%                                           %
% 2. Original System                        %
%                                           %
%%%%%%%%%%%%%%%%%%%%%%%%%%%%%%%%%%%%%%%%%%%%%
\section{Original System}
\label{sec:ori}
\setcounter{equation}{0}

%%%%%%%%%%%%%%%%%%%%%%%%%%%%%%%%%%%%%%%%%%%%%%%%%%%%%%%%%%%%%%%
%                                                             %
%  2.1 General formulation                                    %
%                                                             %
%%%%%%%%%%%%%%%%%%%%%%%%%%%%%%%%%%%%%%%%%%%%%%%%%%%%%%%%%%%%%%%
\subsection{General formulation}
\label{sec:form}

Let us recapitulate the discrete quantum mechanics with real shifts
developed in \cite{os12}.
We restrict ourselves to the finite dimensional matrix case, $x_{\text{max}}=N$.

The Hamiltonian $\mathcal{H}=(\mathcal{H}_{x,y})$ is 
an irreducible (that is, not the direct sum of two or more such matrices) 
tri-diagonal real
symmetric (Jacobi) matrix and its rows and columns are indexed by
non-negative integers $x$ and $y$, $x,y=0,1,\ldots,x_{\text{max}}$.
By adding a scalar matrix to the Hamiltonian, the lowest eigenvalue is
assumed to be zero. This makes the Hamiltonian {\em positive semi-definite}.
By a similarity transformation in terms of a diagonal matrix of
$\pm1$ entries only, the eigenvector corresponding to the zero eigenvalue
can be made to have definite sign, {\em i.e.} all the components are positive
or negative.
Then the Hamiltonian $\mathcal{H}$ has the following form
\begin{equation}
  \mathcal{H}_{x,y}\eqdef
  -\sqrt{B(x)D(x+1)}\,\delta_{x+1,y}-\sqrt{B(x-1)D(x)}\,\delta_{x-1,y}
  +\bigl(B(x)+D(x)\bigr)\delta_{x,y},
  \label{Hdef}
\end{equation}
in which the potential functions $B(x)$ and $D(x)$ are real and positive
but vanish at the boundary:
\begin{align}
  &B(x)>0\ \ (x=0,1,\ldots,x_{\text{max}}-1),\quad B(x_{\text{max}})=0,\n
  &D(x)>0\ \ (x=1,2,\ldots,x_{\text{max}}),\quad D(0)=0.
  \label{BDcondition}
\end{align}
The Schr\"odinger equation is the eigenvalue problem for the hermitian
matrix $\mathcal{H}$,
\begin{equation}
  \mathcal{H}\phi_n(x)=\mathcal{E}_n\phi_n(x)\quad
  (n=0,1,\ldots,n_{\text{max}}),\quad
  0=\mathcal{E}_0<\mathcal{E}_1<\cdots<\mathcal{E}_{n_{\text{max}}},
  \label{schreq0}
\end{equation}
where the eigenvector is $\phi_n=(\phi_n(x))_{x=0,1,\ldots,x_{\text{max}}}$
and $n_{\text{max}}=N$.
Reflecting the {\em positive semi-definiteness} and based on the boundary
conditions \eqref{BDcondition}, the Hamiltonian \eqref{Hdef} can be
expressed in a factorised form:
\begin{align}
  &\mathcal{H}=\mathcal{A}^{\dagger}\mathcal{A},\qquad
  \mathcal{A}=(\mathcal{A}_{x,y}),
  \ \ \mathcal{A}^{\dagger}=((\mathcal{A}^{\dagger})_{x,y})
  =(\mathcal{A}_{y,x}),
  \ \ (x,y=0,1,\ldots,x_{\text{max}}),
  \label{factor}\\
  &\mathcal{A}_{x,y}\eqdef
  \sqrt{B(x)}\,\delta_{x,y}-\sqrt{D(x+1)}\,\delta_{x+1,y},\quad
  (\mathcal{A}^{\dagger})_{x,y}=
  \sqrt{B(x)}\,\delta_{x,y}-\sqrt{D(x)}\,\delta_{x-1,y}.
\end{align}
Here $\mathcal{A}$ ($\mathcal{A}^\dagger$) is an upper (lower) triangular
matrix with the diagonal and the super(sub)-diagonal entries only.
The zero mode equation, $\mathcal{A}\phi_0=0$, is
\begin{align}
  &\sqrt{B(x)}\,\phi_0(x)-\sqrt{D(x+1)}\,\phi_0(x+1)=0
  \ \ (x=0,1,\ldots,x_{\text{max}}-1),\\
  &\sqrt{B(x_{\text{max}})}\,\phi_0(x_{\text{max}})=0,
\end{align}
and the second equation is trivially satisfied by the boundary condition
$B(x_{\text{max}})=0$.
The groundstate eigenvector is easily obtained:
\begin{equation}
  \phi_0(x)=\sqrt{\prod_{y=0}^{x-1}\frac{B(y)}{D(y+1)}}
  \ \ (x=0,1,\ldots,x_{\text{max}}),
  \label{phi0=prodB/D}
\end{equation}
with the normalisation $\phi_0(0)=1$ (convention: $\prod_{k=n}^{n-1}*=1$).
Needless to say it is positive for $x=0,1,\ldots,x_{\text{max}}$.
For the explicit examples treated in \cite{os12}, $\phi_0^2(x)$ can be
analytically continued to the entire complex $x$-plane as a meromorphic
function and it vanishes on the integer points outside the boundary;
$\phi_0^2(x)=0$ ($x\in\mathbb{Z}\backslash\{0,1,\ldots,x_{\text{max}}\}$).
The eigenvectors are mutually orthogonal:
\begin{equation}
  (\phi_n,\phi_m)\eqdef\sum_{x=0}^{x_{\text{max}}}\phi_n(x)\phi_m(x)
  =\frac{1}{d_n^2}\delta_{nm}\quad
  (n,m=0,1,\ldots,n_{\text{max}}).
  \label{ortho}
\end{equation}

For simplicity in notation, we write $\mathcal{H}$, $\mathcal{A}$ and
$\mathcal{A}^{\dagger}$ as follows:
\begin{align}
  &e^{\pm\partial}=((e^{\pm\partial})_{x,y})
  \ \ (x,y=0,1,\ldots,x_{\text{max}}),
  \ \ (e^{\pm\partial})_{x,y}\eqdef\delta_{x\pm 1,y},
  \ \ (e^{\partial})^{\dagger}=e^{-\partial},
  \label{partdef}\\
  &\mathcal{H}=-\sqrt{B(x)}\,e^{\partial}\sqrt{D(x)}
  -\sqrt{D(x)}\,e^{-\partial}\sqrt{B(x)}+B(x)+D(x)\n
  &\phantom{\mathcal{H}}=-\sqrt{B(x)D(x+1)}\,e^{\partial}
  -\sqrt{B(x-1)D(x)}\,e^{-\partial}+B(x)+D(x),
  \label{genham}\\
  &\mathcal{A}=\sqrt{B(x)}-e^{\partial}\sqrt{D(x)},\quad
  \mathcal{A}^{\dagger}=\sqrt{B(x)}-\sqrt{D(x)}\,e^{-\partial}.
  \label{A,Ad}
\end{align}

For the Schr\"odinger equation \eqref{schreq},
it is sufficient that the functions $B(x)$, $D(x)$ and $\phi_n(x)$ are
defined only for the integer grid, $x=0,1,\ldots,x_{\text{max}}$.
In this paper we consider the case that the potential functions $B(x)$
and $D(x)$ are rational functions of $x$ or $q^x$ ($0<q<1$).
So they are defined for any $x\in\mathbb{C}$ (except for the zeros of
their denominators), see the explicit forms \eqref{Bform}--\eqref{Dform}.
Also we consider the eigenvectors in a factorised form:
\begin{equation}
  \phi_n(x)=\phi_0(x)\check{P}_n(x),\quad
  \check{P}_n(x)\eqdef P_n\bigl(\eta(x)\bigr).
  \label{phin=phi0P}
\end{equation}
Here $P_n(\eta)$ is a polynomial of degree $n$ in $\eta$ and the
sinusoidal coordinate $\eta(x)$ is one of the following \cite{os12};
$\eta(x)=x,\epsilon'x(x+d),1-q^x,q^{-x}-1,\epsilon'(q^{-x}-1)(1-dq^x)$,
($\epsilon'=\pm1$).
Since $P_n$ is a polynomial, $\check{P}_n(x)$ is defined for any
$x\in\mathbb{C}$.
The Schr\"odinger equation \eqref{schreq} gives a square root free difference
equation
for the polynomial eigenvector $\check{P}_n(x)$,
\begin{equation}
  B(x)\bigl(\check{P}_n(x)-\check{P}_n(x+1)\bigr)
  +D(x)\bigl(\check{P}_n(x)-\check{P}_n(x-1)\bigr)
  =\mathcal{E}_n\check{P}_n(x)\quad(\forall x\in\mathbb{C}).
  \label{tHcPn=}
\end{equation}

%%%%%%%%%%%%%%%%%%%%%%%%%%%%%%%%%%%%%%%%%%%%%
%                                           %
% 2.2 Original ($q$-)Racah system           %
%                                           %
%%%%%%%%%%%%%%%%%%%%%%%%%%%%%%%%%%%%%%%%%%%%%
\subsection{Original ($q$-)Racah system}
\label{sec:org_qR}

Let us consider the Racah (R) and the $q$-Racah ($q$R) cases.
We follow the notation of \cite{os12}.
Although there are four possible parameter choices indexed by
$(\epsilon,\epsilon')=(\pm 1,\pm 1)$ in general,
as explained in detail in \S\,V.A.1 and \S\,V.A.5 of \cite{os12},
we restrict ourselves
to the $(\epsilon,\epsilon')=(1,1)$ case for simplicity of presentation.
The set of parameters $\bm{\lambda}$, which is different from the
standard one $(\alpha,\beta,\gamma,\delta)$ \cite{koeswart}, its shift
$\bm{\delta}$ and $\kappa$ are
\begin{align}
  \text{R}&:\ \bm{\lambda\,}=(a,b,c,d),\quad \bm{\delta}=(1,1,1,1),
  \quad\kappa=1,
  \label{lamdelR}\\
  \text{$q$R}&:\ q^{\bm{\lambda}}=(a,b,c,d),\quad \bm{\delta}=(1,1,1,1),
  \quad\kappa=q^{-1}, \quad 0<q<1,
  \label{lamdelqR}
\end{align}
where $q^{\bm{\lambda}}$ stands for
$q^{(\lambda_1,\lambda_2,\ldots)}=(q^{\lambda_1},q^{\lambda_2},\ldots)$.
We introduce a new parameter $\tilde{d}$ defined by
\begin{equation}
  \tilde{d}\eqdef
  \left\{
  \begin{array}{ll}
  a+b+c-d-1&:\text{R}\\
  abcd^{-1}q^{-1}&:\text{$q$R}
  \end{array}\right..
\end{equation}
We adopt the following choice of the parameter ranges:
\begin{align}
  \text{R}:\quad&a=-N,\ \ 0<d<a+b,\ \ 0<c<1+d,
  \label{pararange}\\
  \text{$q$R}:\quad&a=q^{-N},\ \ 0<ab<d<1,\ \ qd<c<1,
  \label{pararangeq}
\end{align}
and $x_{\text{max}}=n_{\text{max}}=N$.
They are sufficient for the positivity of $B(x;\bm{\lambda})$ 
and $D(x;\bm{\lambda})$ below.

Here are the fundamental data \cite{os12}:
\begin{align}
  &B(x;\bm{\lambda})=
  \left\{
  \begin{array}{ll}
  {\displaystyle
  -\frac{(x+a)(x+b)(x+c)(x+d)}{(2x+d)(2x+1+d)}}&:\text{R}\\[8pt]
  {\displaystyle-\frac{(1-aq^x)(1-bq^x)(1-cq^x)(1-dq^x)}
  {(1-dq^{2x})(1-dq^{2x+1})}}&:\text{$q$R}
  \end{array}\right.\!,
  \label{Bform}\\
  &D(x;\bm{\lambda})=
  \left\{
  \begin{array}{ll}
  {\displaystyle
  -\frac{(x+d-a)(x+d-b)(x+d-c)x}{(2x-1+d)(2x+d)}}&:\text{R}\\[8pt]
  {\displaystyle-\tilde{d}\,
  \frac{(1-a^{-1}dq^x)(1-b^{-1}dq^x)(1-c^{-1}dq^x)(1-q^x)}
  {(1-dq^{2x-1})(1-dq^{2x})}}&:\text{$q$R}
  \end{array}\right.\!,
  \label{Dform}\\
  &\mathcal{H}(\bm{\lambda})\phi_{n}(x;\bm{\lambda})
  =\mathcal{E}_{n}(\bm{\lambda})
  \phi_{n}(x;\bm{\lambda})
  \ \ (x=0,1,\ldots,x_{\text{max}};n=0,1,\ldots,n_{\text{max}}),
  \label{schreq}\\
  &\phi_{n}(x;\bm{\lambda})=\phi_0(x;\bm{\lambda})\check{P}_n(x;\bm{\lambda}),
  \label{facsol}\\
  &\mathcal{E}_n(\bm{\lambda})=
  \left\{
  \begin{array}{ll}
  n(n+\tilde{d})&:\text{R}\\
  (q^{-n}-1)(1-\tilde{d}q^n)&:\text{$q$R}
  \end{array}\right.\!,\quad
  \eta(x;\bm{\lambda})=
  \left\{
  \begin{array}{ll}
  x(x+d)&:\text{R}\\
  (q^{-x}-1)(1-dq^x)&:\text{$q$R}
  \end{array}\right.\!,
  \label{etadefs}\\
  &
%  f_n(\bm{\lambda})=\mathcal{E}_n(\bm{\lambda}),\quad
%  b_n(\bm{\lambda})=1,\quad
  \varphi(x;\bm{\lambda})=
  \left\{
  \begin{array}{ll}
  {\displaystyle\frac{2x+d+1}{d+1}}&:\text{R}\\[6pt]
  {\displaystyle\frac{q^{-x}-dq^{x+1}}{1-dq}}&:\text{$q$R}
  \end{array}\right.\!,\\
  &\check{P}_n(x;\bm{\lambda})
  =P_n\bigl(\eta(x;\bm{\lambda});\bm{\lambda}\bigr)=
  \left\{
  \begin{array}{ll}
  {\displaystyle
  {}_4F_3\Bigl(
  \genfrac{}{}{0pt}{}{-n,\,n+\tilde{d},\,-x,\,x+d}
  {a,\,b,\,c}\Bigm|1\Bigr)}&:\text{R}\\
  {\displaystyle
  {}_4\phi_3\Bigl(
  \genfrac{}{}{0pt}{}{q^{-n},\,\tilde{d}q^n,\,q^{-x},\,dq^x}
  {a,\,b,\,c}\Bigm|q\,;q\Bigr)}&:\text{$q$R}
  \end{array}\right.
  \label{qracah}\\
  &\phantom{\check{P}_n(x;\bm{\lambda})
  =P_n\bigl(\eta(x;\bm{\lambda});\bm{\lambda}\bigr)}
  =\left\{
  \begin{array}{ll}
  {\displaystyle
  R_n\bigl(\eta(x;\bm{\lambda});a-1,\tilde{d}-a,c-1,d-c\bigr)}
  &:\text{R}\\
  {\displaystyle
  R_n\bigl(1+d+\eta(x;\bm{\lambda});
  aq^{-1},\tilde{d}a^{-1},cq^{-1},dc^{-1}|q\bigr)}&:\text{$q$R}
  \end{array}\right.\!,
  \label{Pn=R,qR}\\
  &\phi_0(x;\bm{\lambda})^2=\left\{
  \begin{array}{ll}
  {\displaystyle
  \frac{(a,b,c,d)_x}{(1+d-a,1+d-b,1+d-c,1)_x}\,
  \frac{2x+d}{d}
  }&:\text{R}\\[8pt]
  {\displaystyle
  \frac{(a,b,c,d\,;q)_x}
  {(a^{-1}dq,b^{-1}dq,c^{-1}dq,q\,;q)_x\,\tilde{d}^x}\,
  \frac{1-dq^{2x}}{1-d}
  }&:\text{$q$R}
  \end{array}\right.\!,\\
  &d_n(\bm{\lambda})^2
  =\left\{
  \begin{array}{ll}
  {\displaystyle
  \frac{(a,b,c,\tilde{d})_n}
  {(1+\tilde{d}-a,1+\tilde{d}-b,1+\tilde{d}-c,1)_n}\,
  \frac{2n+\tilde{d}}{\tilde{d}}
  }&\\[8pt]
  {\displaystyle
  \quad\times
  \frac{(-1)^N(1+d-a,1+d-b,1+d-c)_N}{(\tilde{d}+1)_N(d+1)_{2N}}
  }&:\text{R}\\[8pt]
  {\displaystyle
  \frac{(a,b,c,\tilde{d}\,;q)_n}
  {(a^{-1}\tilde{d}q,b^{-1}\tilde{d}q,c^{-1}\tilde{d}q,q\,;q)_n\,d^n}\,
  \frac{1-\tilde{d}q^{2n}}{1-\tilde{d}}
  }&\\[8pt]
  {\displaystyle
  \quad\times
  \frac{(-1)^N(a^{-1}dq,b^{-1}dq,c^{-1}dq\,;q)_N\,\tilde{d}^Nq^{\frac12N(N+1)}}
  {(\tilde{d}q\,;q)_N(dq\,;q)_{2N}}
  }&:\text{$q$R}
  \end{array}\right.\!.
  \label{dn2}
\end{align}
Here $R_n(\cdots)$ in \eqref{Pn=R,qR} are the standard notation of the
($q$-)Racah polynomial in \cite{koeswart}.
It should be emphasised that the quantities $B(x;\bm{\lambda})$,
$D(x;\bm{\lambda})$, $\mathcal{E}_n(\bm{\lambda})$,
$\check{P}_n(x;\bm{\lambda})$, $\phi_0(x;\bm{\lambda})^2$,
$d_n(\bm{\lambda})^2$ are formally symmetric under the permutation
of $(a,b,c)$, although their ranges are restricted as above by
\eqref{pararange}--\eqref{pararangeq}.

Here is a remark on the polynomial $\check{P}_n(x;\bm{\lambda})$,
which is in fact a polynomial in the sinusoidal coordinate
$\eta(x;\bm{\lambda})$ \eqref{etadefs}.
The sinusoidal coordinate has a special dynamical meaning \cite{os12,os13,os7}.
The Heisenberg operator solution for $\eta(x;\bm{\lambda})$ can be expressed
in a closed form. This means that its time evolution is a sinusoidal motion.
Let $R$ be the ring of polynomials in $x$ (the Racah case) or the ring
of Laurent polynomials in $q^x$ (the $q$-Racah case). Let us introduce an
automorphism $\mathcal{I}$ in $R$ by
\begin{equation}
  \mathcal{I}(x)=-x-d \quad :\text{R},\qquad
  \mathcal{I}(q^x)=q^{-x}d^{-1} \quad :q\text{R}.
  \label{autom}
\end{equation}
Obviously it is an involution $\mathcal{I}^2=\text{id}$.
The following remark is important.\\
{\bf Remark}: If a (Laurent) polynomial $\check{f}$ in $x$ ($q^x$)
is invariant under the above involution, it is a polynomial in the
sinusoidal coordinate $\eta(x;\bm{\lambda})$:
\begin{equation}
  \mathcal{I}\bigl(\check{f}(x)\bigr)=\check{f}(x)
  \ \Leftrightarrow\ \check{f}(x)=f\bigl(\eta(x;\bm{\lambda})\bigr).
  \label{remark}
\end{equation}

The system is shape invariant \cite{genden,os12},
\begin{equation}
  \mathcal{A}(\bm{\lambda})\mathcal{A}(\bm{\lambda})^{\dagger}
  =\kappa\mathcal{A}(\bm{\lambda}+\bm{\delta})^{\dagger}
  \mathcal{A}(\bm{\lambda}+\bm{\delta})+\mathcal{E}_1(\bm{\lambda}),
\end{equation}
which is a sufficient condition for exact solvability and it provides
the explicit formulas for the energy eigenvalues and the eigenfunctions,
{\em i.e.} the generalised Rodrigues formula \cite{os12}.
The forward and backward shift relations are
\begin{equation}
  \mathcal{F}(\bm{\lambda})\check{P}_n(x;\bm{\lambda})
  =\mathcal{E}_n(\bm{\lambda})\check{P}_{n-1}(x;\bm{\lambda}+\bm{\delta}),
  \quad
%  \label{FPn=EnPn-1}\\
  \mathcal{B}(\bm{\lambda})\check{P}_{n-1}(x;\bm{\lambda}+\bm{\delta})
  =\check{P}_n(x;\bm{\lambda}),
  \label{BPn-1=Pn}
\end{equation}
where the forward and backward shift operators are
\begin{equation}
  \mathcal{F}(\bm{\lambda})=B(0;\bm{\lambda})\varphi(x;\bm{\lambda})^{-1}
  (1-e^{\partial}),
  \ \ \mathcal{B}(\bm{\lambda})=B(0;\bm{\lambda})^{-1}
  \bigl(B(x;\bm{\lambda})-D(x;\bm{\lambda})e^{-\partial}\bigr)
  \varphi(x;\bm{\lambda}).
\end{equation}

%%%%%%%%%%%%%%%%%%%%%%%%%%%%%%%%%%%%%%%%%%%%%%%
%                                             %
% 2.3 Symmetry and virtual state vectors      %
%                                             %
%%%%%%%%%%%%%%%%%%%%%%%%%%%%%%%%%%%%%%%%%%%%%%%
\subsection{Symmetry and virtual state vectors}
\label{sec:virtual}

Let us define the twist operation $\mathfrak{t}$ of the parameters:
\begin{equation}
  \mathfrak{t}(\bm{\lambda})\eqdef
  (\lambda_4-\lambda_1+1,\lambda_4-\lambda_2+1,\lambda_3,\lambda_4),
  \quad \mathfrak{t}^2=\text{id}.
  \label{twist}
\end{equation}
We introduce two functions $B'(x)$ and $D'(x)$  by
\begin{equation}
  B'(x;\bm{\lambda})\eqdef B\bigl(x;\mathfrak{t}(\bm{\lambda})\bigr),\quad
  D'(x;\bm{\lambda})\eqdef D\bigl(x;\mathfrak{t}(\bm{\lambda})\bigr),
  \label{B'D'}
\end{equation}
namely,
\begin{align}
  &B'(x;\bm{\lambda})=
  \left\{
  \begin{array}{ll}
  {\displaystyle
  -\frac{(x+d-a+1)(x+d-b+1)(x+c)(x+d)}{(2x+d)(2x+1+d)}}&:\text{R}\\[8pt]
  {\displaystyle
  -\frac{(1-a^{-1}dq^{x+1})(1-b^{-1}dq^{x+1})(1-cq^x)(1-dq^x)}
  {(1-dq^{2x})(1-dq^{2x+1})}}&:\text{$q$R}
  \end{array}\right.\!,\\
  &D'(x;\bm{\lambda})=
  \left\{
  \begin{array}{ll}
  {\displaystyle
  -\frac{(x+a-1)(x+b-1)(x+d-c)x}{(2x-1+d)(2x+d)}}&:\text{R}\\[8pt]
  {\displaystyle-\frac{cdq}{ab}\,
  \frac{(1-aq^{x-1})(1-bq^{x-1})(1-c^{-1}dq^x)(1-q^x)}
  {(1-dq^{2x-1})(1-dq^{2x})}}&:\text{$q$R}
  \end{array}\right.\!.
\end{align}
We restrict the parameter range
\begin{equation}
  \text{R}:\ \ d+M<a+b,\qquad
  \text{$q$R}:\ \ ab<dq^M,
  \label{Mrange}
\end{equation}
in which $M$ is a positive integer and later it will be identified with
the total number of deleted virtual states.
It is easy to verify
\begin{align}
  &B(x;\bm{\lambda})D(x+1;\bm{\lambda})
  =\alpha(\bm{\lambda})^2B'(x;\bm{\lambda})D'(x+1;\bm{\lambda}),
  \label{BD=B'D'}\\
  &B(x;\bm{\lambda})+D(x;\bm{\lambda})
  =\alpha(\bm{\lambda})\bigl(B'(x;\bm{\lambda})
  +D'(x;\bm{\lambda})\bigr)+\alpha'(\bm{\lambda}),
  \label{BD=B'D'2}\\
  &B'(x;\bm{\lambda})>0\ \ (x=0,1,\ldots,x_{\text{max}}+M-1),
  \label{B'>0,..}\\
  &D'(x;\bm{\lambda})>0\ \ (x=1,2,\ldots,x_{\text{max}}),
  \ \ D'(0;\bm{\lambda})=D'(x_{\text{max}}+1;\bm{\lambda})=0.
  \label{D'>0,..}
\end{align}
Here the constant $\alpha(\bm{\lambda})$ is positive and
$\alpha'(\bm{\lambda})$ is negative:
\begin{align}
  &0<\alpha(\bm{\lambda})=\left\{
  \begin{array}{ll}
  1&:\text{R}\\
  abd^{-1}q^{-1}&:\text{$q$R}
  \end{array}\right.,\quad
  0>\alpha'(\bm{\lambda})=\left\{
  \begin{array}{ll}
  -c(a+b-d-1)&:\text{R}\\
  -(1-c)(1-abd^{-1}q^{-1})&:\text{$q$R}
  \end{array}\right..
\end{align}
The above relations \eqref{BD=B'D'}--\eqref{D'>0,..} imply that we can
define a virtual Hamiltonian $\mathcal{H}'$ by the twisted parameters 
(the $\bm{\lambda}$ dependence is suppressed for simplicity):
\begin{align}
  \mathcal{H}'(\bm{\lambda})&\eqdef
  \mathcal{H}\bigl(\mathfrak{t}(\bm{\lambda})\bigr)
  =-\sqrt{B'(x)}\,e^{\partial}\sqrt{D'(x)}
  -\sqrt{D'(x)}\,e^{-\partial}\sqrt{B'(x)}+B'(x)+D'(x),
  \label{Hprime}
\end{align}
and the original Hamiltonian and the virtual Hamiltonian are linearly 
related 
\begin{align}
  &\mathcal{H}(\bm{\bm{\lambda}})=
  \alpha(\bm{\lambda})\mathcal{H}\bigl(\mathfrak{t}(\bm{\lambda})\bigr)
  +\alpha'(\bm{\lambda}).
  \label{HH'}
  \end{align}
This also means that $\mathcal{H}(\mathfrak{t}(\bm{\lambda}))$
is {\em positive definite} and it has {\em no zero-mode}.
In other words, the two term recurrence relation determining
the `zero-mode' of $\mathcal{H}(\mathfrak{t}(\bm{\lambda}))$
\begin{align}
  &\mathcal{A}\bigl(\mathfrak{t}(\bm{\lambda})\bigr)=
  \sqrt{B'(x;\bm{\lambda})}-e^{\partial}\sqrt{D'(x;\bm{\lambda})},\n
  &\mathcal{A}\bigl(\mathfrak{t}(\bm{\lambda})\bigr)
  \tilde{\phi}_0(x;\bm{\lambda})=0\quad(x=0,1,\ldots,x_{\text{max}}-1),
  \label{aprimezero}
\end{align}
can be `solved' from $x=0$ to $x=x_\text{max}-1$ to determine
\begin{equation}
  \tilde{\phi}_0(x;\bm{\lambda})\eqdef
  \sqrt{\prod_{y=0}^{x-1}\frac{B'(y;\bm{\lambda})}{D'(y+1;\bm{\lambda})}}
  \qquad (x=0,1,\ldots,x_{\text{max}}).
  \label{tphi0}
\end{equation}
But at the end point $x=x_\text{max}$, the `zero-mode' equation
\eqref{aprimezero} is not satisfied, because of the boundary condition
\begin{equation}
  B'(x_\text{max};\bm{\lambda})\neq 0,\quad
  \mathcal{A}\bigl(\mathfrak{t}(\bm{\lambda})\bigr)
  \tilde{\phi}_0(x;\bm{\lambda})\neq0\ \ (x=x_{\text{max}}).
  \label{virtnon}
\end{equation}
The new Schr\"odinger equation
\begin{equation}
  \mathcal{H}'(\bm{\lambda})\tilde{\phi}_{\text{v}}(x;\bm{\lambda})
  =\mathcal{E}'_{\text{v}}(\bm{\lambda})
  \tilde{\phi}_{\text{v}}(x;\bm{\lambda}),
  \label{schreq'}
\end{equation}
can be {\em almost solved} except for the end point $x=x_\text{max}$ by
the factorisation ansatz
\begin{equation*}
  \tilde{\phi}_{\text{v}}(x;\bm{\lambda})
  \eqdef\tilde{\phi}_0(x;\bm{\lambda})
  \check{\xi}_{\text{v}}(x;\bm{\lambda}),
\end{equation*}
as in the original ($q$-)Racah system.
By using the explicit form of $\tilde{\phi}_0(x;\bm{\lambda})$
\eqref{tphi0}, the new Schr\"odinger equation for
$x=0,\ldots, x_\text{max}-1$ is rewritten as
\begin{equation}
  B'(x;\bm{\lambda})\bigl(\check{\xi}_{\text{v}}(x;\bm{\lambda})
  -\check{\xi}_{\text{v}}(x+1;\bm{\lambda})\bigr)
  +D'(x;\bm{\lambda})\bigl(\check{\xi}_{\text{v}}(x;\bm{\lambda})
  -\check{\xi}_{\text{v}}(x-1;\bm{\lambda})\bigr)
  =\mathcal{E}'_{\text{v}}(\bm{\lambda})\check{\xi}_{\text{v}}(x;\bm{\lambda}).
  \label{tH'cxi=}
\end{equation}
This is the same form of equation as that for the ($q$-)Racah polynomials.
So its solution for $x\in\mathbb{C}$ is given by the
($q$-)Racah polynomial \eqref{qracah} with the twisted parameters:
\begin{equation}
  \check{\xi}_{\text{v}}(x;\bm{\lambda})
  =\check{P}_{\text{v}}\bigl(x;\mathfrak{t}(\bm{\lambda})\bigr),\qquad
  \mathcal{E}'_{\text{v}}(\bm{\lambda})
  =\mathcal{E}_{\text{v}}\bigl(\mathfrak{t}(\bm{\lambda})\bigr).
\end{equation}
Among such `solutions', those with the negative energy and having 
definite sign
\begin{align}
  \check{\xi}_{\text{v}}(x;\bm{\lambda})>0&
  \ \ (x=0,1,\ldots,x_{\text{max}},x_{\text{max}}+1;
  \text{v}\in\mathcal{V}),
  \label{xi>0}\\
  \tilde{\mathcal{E}}_{\text{v}}(\bm{\lambda})<0&\ \ (\text{v}\in\mathcal{V}),
  \label{tEv<0}
\end{align}
are called the {\em virtual state vectors\/}:
 $\{\tilde{\phi}_{\text{v}}(x)\}$, $\text{v}\in \mathcal{V}$.
The index set of the virtual state vectors is
\begin{equation}
  \mathcal{V}=\{1,2,\ldots,\text{v}_{\text{max}}\},
  \quad
  \text{v}_{\text{max}}=\min\bigl\{
  [\lambda_1+\lambda_2-\lambda_4-1]',
  [\tfrac12(\lambda_1+\lambda_2-\lambda_3-\lambda_4)]\bigr\},
  \label{vrange}
\end{equation}
where $[x]$ denotes the greatest integer not exceeding $x$ and
$[x]'$ denotes the greatest integer not equal or exceeding $x$.
We will not use the label 0 state for deletion, see \eqref{dM=0}.
The negative virtual state energy conditions \eqref{tEv<0} is met by
$\text{v}_{\text{max}}\leq[\lambda_1+\lambda_2-\lambda_4-1]'$.
For the positivity of $\check{\xi}_{\text{v}}(x;\bm{\lambda})$ \eqref{xi>0},
we write down them explicitly:
\begin{align}
  \check{\xi}_{\text{v}}(x;\bm{\lambda})&=
  \left\{
  \begin{array}{ll}
  {\displaystyle
  {}_4F_3\Bigl(
  \genfrac{}{}{0pt}{}{-\text{v},\,\text{v}-a-b+c+d+1,\,-x,\,x+d}
  {d-a+1,\,d-b+1,\,c}\Bigm|1\Bigr)}&:\text{R}\\[8pt]
  {\displaystyle
  {}_4\phi_3\Bigl(
  \genfrac{}{}{0pt}{}{q^{-\text{v}},\,a^{-1}b^{-1}cdq^{\text{v}+1},
  \,q^{-x},\,dq^x}
  {a^{-1}dq,\,b^{-1}dq,\,c}\Bigm|q\,;q\Bigr)}&:\text{$q$R}
  \end{array}\right.\n
  &=\left\{
  \begin{array}{ll}
  {\displaystyle
  \sum_{k=0}^{\text{v}}\frac{(-\text{v},\text{v}-a-b+c+d+1,-x,x+d)_k}
  {(d-a+1,d-b+1,c)_k}\frac{1}{k!}}&:\text{R}\\[8pt]
  {\displaystyle
  \sum_{k=0}^{\text{v}}\frac{(q^{-\text{v}},a^{-1}b^{-1}cdq^{\text{v}+1},
  q^{-x},dq^x;q)_k}
  {(a^{-1}dq,b^{-1}dq,c;q)_k}\frac{q^k}{(q;q)_k}}&:\text{$q$R}
  \end{array}\right..
  \label{xipos}
\end{align}
Each $k$-th term in the sum is non-negative
for $2\text{v}_{\text{max}}\leq\lambda_1+\lambda_2-\lambda_3-\lambda_4$.
 
Here is a summary of the properties of the virtual state
vectors:
\begin{align}
    &\tilde{\phi}_0(x;\bm{\lambda})
  \eqdef\phi_0\bigl(x;\mathfrak{t}(\bm{\lambda})\bigr),\quad
  \tilde{\phi}_{\text{v}}(x;\bm{\lambda})
  \eqdef\phi_{\text{v}}\bigl(x;\mathfrak{t}(\bm{\lambda})\bigr)
  =\tilde{\phi}_0(x;\bm{\lambda})
  \check{\xi}_{\text{v}}(x;\bm{\lambda})
  \ \ (\text{v}\in\mathcal{V}),
  \label{tphiv=}\\
  &\check{\xi}_{\text{v}}(x;\bm{\lambda})\eqdef
  \check{P}_{\text{v}}\bigl(x;\mathfrak{t}(\bm{\lambda})\bigr),\quad
  \check{\xi}_{\text{v}}(x;\bm{\lambda})\eqdef
  \xi_{\text{v}}\bigl(\eta(x;\bm{\lambda});\bm{\lambda}\bigr),
  \label{xiv=}\\
  &\mathcal{H}(\bm{\lambda})\tilde{\phi}_{\text{v}}(x;\bm{\lambda})
  =\tilde{\mathcal{E}}_{\text{v}}(\bm{\lambda})
  \tilde{\phi}_{\text{v}}(x;\bm{\lambda})
  \ \ (x=0,1,\ldots,x_{\text{max}}-1),\n
  &\mathcal{H}(\bm{\lambda})
  \tilde{\phi}_{\text{v}}(x_{\text{max}};\bm{\lambda})
  \neq\tilde{\mathcal{E}}_{\text{v}}(\bm{\lambda})
  \tilde{\phi}_{\text{v}}(x_{\text{max}};\bm{\lambda}),
  \qquad\quad
  \mathcal{E}'_{\text{v}}(\bm{\lambda})
  =\mathcal{E}_{\text{v}}\bigl(\mathfrak{t}(\bm{\lambda})\bigr),
  \label{H'tphiv=}\\
  &\tilde{\mathcal{E}}_{\text{v}}(\bm{\lambda})
  =\alpha(\bm{\lambda})\mathcal{E}'_{\text{v}}(\bm{\lambda})
  +\alpha'(\bm{\lambda})
  =\left\{
  \begin{array}{ll}
  -(c+\text{v})(a+b-d-1-\text{v})&:\text{R}\\
  -(1-cq^{\text{v}})(1-abd^{-1}q^{-1-\text{v}})&:\text{$q$R}
  \end{array}\right.,
  \label{tEv=}\\
  &\nu(x;\bm{\lambda})\eqdef
  \frac{\phi_0(x;\bm{\lambda})}{\tilde{\phi}_0(x;\bm{\lambda})}
  =\left\{
  \begin{array}{ll}
  {\displaystyle
  \frac{\Gamma(1-a)\Gamma(x+b)\Gamma(d-a+1)\Gamma(b-d-x)}
  {\Gamma(1-a-x)\Gamma(b)\Gamma(x+d-a+1)\Gamma(b-d)}}
  &:\text{R}\\[8pt]
  {\displaystyle
  \frac{(a^{-1}q^{1-x},b,a^{-1}dq^{x+1},bd^{-1};q)_{\infty}}
  {(a^{-1}q,bq^x,a^{-1}dq,bd^{-1}q^{-x};q)_{\infty}}}
  &:\text{$q$R}
  \end{array}\right..
  \label{phi0/tphi0}
\end{align}
Note that $\alpha'(\bm{\lambda})=\tilde{\mathcal{E}}_0(\bm{\lambda})<0$.
The function $\nu(x;\bm{\lambda})$ can be analytically continued into a
meromorphic function of $x$ or $q^x$ through the functional relations:
\begin{equation}
  \nu(x+1;\bm{\lambda})=\frac{B(x;\bm{\lambda})}{\alpha B'(x;\bm{\lambda})}
  \nu(x;\bm{\lambda}),\quad
  \nu(x-1;\bm{\lambda})=\frac{D(x;\bm{\lambda})}{\alpha D'(x;\bm{\lambda})}
  \nu(x;\bm{\lambda}).
  \label{nurel}
\end{equation}
By $B(x_{\text{max}};\bm{\lambda})=0$, it vanishes for integer
$x_\text{max}+1\leq x\leq x_{\text{max}}+M$, $\nu(x;\bm{\lambda})=0$,
and at negative integer points it takes nonzero finite values in general.

%%%%%%%%%%%%%%%%%%%%%%%%%%%%%%%%%%%%%%%%%%%%%%%%%%%%%%%%%%%%%%%
%                                                             %
%  3. Multi-indexed ($q$-)Racah Polynomials                   %
%                                                             %
%%%%%%%%%%%%%%%%%%%%%%%%%%%%%%%%%%%%%%%%%%%%%%%%%%%%%%%%%%%%%%%
\section{Multi-indexed ($q$-)Racah Polynomials}
\label{sec:miop_qR}
\setcounter{equation}{0}

In this section we apply the Crum-Adler method of virtual states deletion to
the exactly solvable systems whose eigenstates are described by
the ($q$-)Racah polynomials.
Since all the eigenvalues remain the same, {\em i.e.\/}
the process is {\em exactly iso-spectral deformation\/},
the size of the Hamiltonian is unchanged.

Various quantities are neatly expressed in terms of a Casoratian,
a discrete counterpart of the Wronskian.
The Casorati determinant of a set of $n$ functions $\{f_j(x)\}$ is defined by
\begin{equation*}
  \text{W}[f_1,\ldots,f_n](x)
  \eqdef\det\Bigl(f_k(x+j-1)\Bigr)_{1\leq j,k\leq n},
\end{equation*}
(for $n=0$, we set $\text{W}[\cdot](x)=1$), which satisfies identities
\begin{align*}
  &\text{W}[gf_1,gf_2,\ldots,gf_n](x)
  =\prod_{k=0}^{n-1}g(x+k)\cdot\text{W}[f_1,f_2,\ldots,f_n](x),
%  \label{Wformula1}
  \\
  &\text{W}\bigl[\text{W}[f_1,f_2,\ldots,f_n,g],
  \text{W}[f_1,f_2,\ldots,f_n,h]\,\bigr](x)\n
  &=\text{W}[f_1,f_2,\ldots,f_n](x+1)\,
  \text{W}[f_1,f_2,\ldots,f_n,g,h](x)
  \quad(n\geq 0).
%  \label{Wformula2}
\end{align*}

%%%%%%%%%%%%%%%%%%%%%%%%%%%%%%%%%%%%%%%%%%%%%%%
%                                             %
% 3.1 Virtual states deletion                 %
%                                             %
%%%%%%%%%%%%%%%%%%%%%%%%%%%%%%%%%%%%%%%%%%%%%%%
\subsection{Virtual states deletion}
\label{sec:virtual_del}

Let us provide the basic formulas starting from  one virtual
state deletion. For simplicity of presentation the parameter 
($\bm{\lambda}$) dependence of various quantities is suppressed in
this subsection.

\medskip

\noindent
\underline{one virtual state vector deletion}

First we rewrite the original Hamiltonian by introducing potential functions
$\hat{B}_{d_1}(x)$ and $\hat{D}_{d_1}(x)$ determined by one of the virtual
state polynomials $\check{\xi}_{d_1}(x)$ ($d_1\in\mathcal{V}$):
\begin{equation}
  \hat{B}_{d_1}(x)\eqdef\alpha B'(x)
  \frac{\check{\xi}_{d_1}(x+1)}{\check{\xi}_{d_1}(x)},\quad
  \hat{D}_{d_1}(x)\eqdef\alpha D'(x)
  \frac{\check{\xi}_{d_1}(x-1)}{\check{\xi}_{d_1}(x)}.
  \label{Bd1def}
\end{equation}
We have $\hat{B}_{d_1}(x)>0$ ($x=0,1,\ldots,x_{\text{max}}$),
$\hat{D}_{d_1}(0)=\hat{D}_{d_1}(x_{\text{max}}+1)=0$,
$\hat{D}_{d_1}(x)>0$ ($x=1,2,\ldots,x_{\text{max}}$) and
\begin{align*}
  &B(x)D(x+1)=\hat{B}_{d_1}(x)\hat{D}_{d_1}(x+1),\\
  &B(x)+D(x)=\hat{B}_{d_1}(x)+\hat{D}_{d_1}(x)+\tilde{\mathcal{E}}_{d_1},
\end{align*}
where use is made of \eqref{tH'cxi=} in the second equation.
The original Hamiltonian reads:
\begin{align*}
  &\mathcal{H}=\hat{\mathcal{A}}_{d_1}^{\dagger}\hat{\mathcal{A}}_{d_1}
  +\tilde{\mathcal{E}}_{d_1},\\
  &\hat{\mathcal{A}}_{d_1}\eqdef
  \sqrt{\hat{B}_{d_1}(x)}-e^{\partial}\sqrt{\hat{D}_{d_1}(x)},\quad
  \hat{\mathcal{A}}_{d_1}^{\dagger}
  =\sqrt{\hat{B}_{d_1}(x)}-\sqrt{\hat{D}_{d_1}(x)}\,e^{-\partial}.
\end{align*}
The virtual state vector $\tilde{\phi}_{d_1}(x)$ is almost annihilated
by $\hat{\mathcal{A}}_{d_1}$, except for the upper end point:
\begin{equation*}
  \hat{\mathcal{A}}_{d_1}\tilde{\phi}_{d_1}(x)=0
  \ \ (x=0,1,\ldots,x_{\text{max}}-1),\quad
  \hat{\mathcal{A}}_{d_1}\tilde{\phi}_{d_1}(x_{\text{max}})\neq 0.
\end{equation*}
The proof is straightforward by direct substitution of 
\eqref{Bd1def} and \eqref{tphi0}.

Next let us define a new Hamiltonian $\mathcal{H}_{d_1}$ by changing
the order of the two matrices $\hat{\mathcal{A}}_{d_1}^{\dagger}$ and
$\hat{\mathcal{A}}_{d_1}$ together with the sets of new eigenvectors
$\phi_{d_1n}(x)$ and new virtual state vectors
$\tilde{\phi}_{d_1\text{v}}(x)$:
\begin{align}
  \mathcal{H}_{d_1}&\eqdef
  \hat{\mathcal{A}}_{d_1}\hat{\mathcal{A}}_{d_1}^{\dagger}
  +\tilde{\mathcal{E}}_{d_1},\quad
  \mathcal{H}_{d_1}=(\mathcal{H}_{d_1\,x,y})
  \ \ (x,y=0,1,\ldots,x_{\text{max}}),\\
  \phi_{d_1n}(x)&\eqdef\hat{\mathcal{A}}_{d_1}\phi_n(x)
  \ \ (x=0,1,\ldots,x_{\text{max}}; n=0,1,\ldots,n_{\text{max}}),
  \label{phid1n}\\
  \tilde{\phi}_{d_1\text{v}}(x)&\eqdef
  \hat{\mathcal{A}}_{d_1}\tilde{\phi}_{\text{v}}(x)
  +\delta_{x,x_{\text{max}}}\varphi_{d_1\text{v}}
  \ \ (x=0,1,\ldots,x_{\text{max}};\text{v}\in\mathcal{V}\backslash\{d_1\}),
  \n
  &\varphi_{d_1\text{v}}\eqdef
  -\frac{\sqrt{\alpha B'(x_{\text{max}})}\,\tilde{\phi}_0(x_{\text{max}})}
  {\sqrt{\check{\xi}_{d_1}(x_{\text{max}})\check{\xi}_{d_1}(x_{\text{max}}+1)}}
  \,\check{\xi}_{d_1}(x_{\text{max}})\check{\xi}_{\text{v}}(x_{\text{max}}+1).
  \label{tphid1v}
\end{align}
The $\varphi_{d_1\text{v}}$ term is necessary for the Casoratian
expression for $\tilde{\phi}_{d_1\text{v}}(x)$ in \eqref{phid1v} to hold
at $x=x_{\text{max}}$.
It is easy to verify that $\phi_{d_1n}(x)$ is an eigenvector and that
$\tilde{\phi}_{d_1\text{v}}(x)$ is a virtual state vector
\begin{align*}
  &\mathcal{H}_{d_1}\phi_{d_1n}(x)
  =\mathcal{E}_n\phi_{d_1n}(x)
  \ \ (x=0,1,\ldots,x_{\text{max}}; n=0,1,\ldots,n_{\text{max}}),\\
  &\mathcal{H}_{d_1}\tilde{\phi}_{d_1\text{v}}(x)
  =\tilde{\mathcal{E}}_{\text{v}}\tilde{\phi}_{d_1\text{v}}(x)
  \ \ (x=0,1,\ldots,x_{\text{max}}-1;
  \text{v}\in\mathcal{V}\backslash\{d_1\}),\n
  &\mathcal{H}_{d_1}\tilde{\phi}_{d_1\text{v}}(x_{\text{max}})
  \neq\tilde{\mathcal{E}}_{\text{v}}
  \tilde{\phi}_{d_1\text{v}}(x_{\text{max}}).
%  \label{virtI}
\end{align*}
For example,
\begin{align*}
  &\mathcal{H}_{d_1}\phi_{d_1n}
  =(\hat{\mathcal{A}}_{d_1}\hat{\mathcal{A}}_{d_1}^{\dagger}
  +\tilde{\mathcal{E}}_{d_1})\hat{\mathcal{A}}_{d_1}\phi_n
  =\hat{\mathcal{A}}_{d_1}(\hat{\mathcal{A}}_{d_1}^{\dagger}
  \hat{\mathcal{A}}_{d_1}+\tilde{\mathcal{E}}_{d_1})\phi_n\\
  &=\hat{\mathcal{A}}_{d_1}\mathcal{H}\phi_n
  =\hat{\mathcal{A}}_{d_1}\mathcal{E}_n\phi_n
  =\mathcal{E}_n\hat{\mathcal{A}}_{d_1}\phi_n
  =\mathcal{E}_n\phi_{d_1n}.
\end{align*}
The two Hamiltonians $\mathcal{H}$ and $\mathcal{H}_{d_1}$ are exactly
iso-spectral. If the original system is exactly solvable, this new system
is also exactly solvable.
The orthogonality relation for the new eigenvectors is
\begin{align*}
  &\quad(\phi_{d_1n},\phi_{d_1m})
  =\sum_{x=0}^{x_{\text{max}}}\phi_{d_1n}(x)\phi_{d_1m}(x)\n
  &=(\hat{\mathcal{A}}_{d_1}\phi_n,\hat{\mathcal{A}}_{d_1}\phi_m)
  =(\hat{\mathcal{A}}_{d_1}^{\dagger}\hat{\mathcal{A}}_{d_1}\phi_n,\phi_m)
  =\bigl((\mathcal{H}-\tilde{\mathcal{E}}_{d_1})\phi_n,\phi_m\bigr)\n
  &=(\mathcal{E}_n-\tilde{\mathcal{E}}_{d_1})(\phi_n,\phi_m)
  =(\mathcal{E}_n-\tilde{\mathcal{E}}_{d_1})\frac{1}{d_n^2}\delta_{nm}
  \ \ (n,m=0,1,\ldots,n_{\text{max}}).
\end{align*}
This shows clearly that the {\em negative} virtual state energy
($\tilde{\mathcal{E}}_{\text{v}}<0$) is necessary for the positivity
of the inner products.

The new eigenvector $\phi_{d_1n}(x)$ \eqref{phid1n} and the virtual state
vector $\tilde{\phi}_{d_1\text{v}}(x)$ \eqref{tphid1v} are expressed neatly
in terms of the Casoratian ($x=0,1,\ldots,x_{\text{max}}$)
\begin{equation}
  \phi_{d_1n}(x)=\frac{-\sqrt{\alpha B'(x)}\,\tilde{\phi}_0(x)}
  {\sqrt{\check{\xi}_{d_1}(x)\check{\xi}_{d_1}(x+1)}}\,
  \text{W}\bigl[\check{\xi}_{d_1},\nu\check{P}_n\bigr](x),
  \ \ \tilde{\phi}_{d_1\text{v}}(x)
  =\frac{-\sqrt{\alpha B'(x)}\,\tilde{\phi}_0(x)}
  {\sqrt{\check{\xi}_{d_1}(x)\check{\xi}_{d_1}(x+1)}}\,
  \text{W}[\check{\xi}_{d_1},\check{\xi}_{\text{v}}](x).
  \label{phid1v}
\end{equation}
We will show that
the positivity of the virtual state vector is inherited by the new virtual
state vector $\tilde{\phi}_{d_1\text{v}}(x)$ \eqref{tphid1v}.
The Casoratian  $\text{W}[\check{\xi}_{d_1},\check{\xi}_{\text{v}}](x)$
has definite sign for $x=0,1,\ldots,x_{\text{max}}+1$, namely all positive
or all negative.
By using \eqref{tH'cxi=} we have
\begin{equation*}
  \alpha B'(x)\text{W}[\check{\xi}_{d_1},\check{\xi}_{\text{v}}](x)
  =\alpha D'(x)\text{W}[\check{\xi}_{d_1},\check{\xi}_{\text{v}}](x-1)
  +(\tilde{\mathcal{E}}_{d_1}-\tilde{\mathcal{E}}_{\text{v}})
  \check{\xi}_{d_1}(x)\check{\xi}_{\text{v}}(x).
%  \label{B'W=}
\end{equation*}
By setting $x=0,1,\ldots,x_{\text{max}}+1$ in turn, we obtain
\begin{equation*}
  \pm(\tilde{\mathcal{E}}_{d_1}-\tilde{\mathcal{E}}_{\text{v}})>0
  \Rightarrow
  \pm\text{W}[\check{\xi}_{d_1},\check{\xi}_{\text{v}}](x)>0
  \ \ (x=0,1,\ldots,x_{\text{max}}+1).
\end{equation*}
Note that the set of virtual eigenvalues
$\{\tilde{\mathcal{E}}_{d_j}\}$ are mutually distinct.
We will now show that
the new groundstate eigenvector $\phi_{d_10}(x)$ is of
definite sign as the original one $\phi_0(x)$
\eqref{phi0=prodB/D}.
We show that the Casoratian $\text{W}[\check{\xi}_{d_1},\nu](x)$ has definite
sign for $x=0,1,\ldots,x_{\text{max}}$.
By writing down the equation $\mathcal{H}\phi_n(x)=\mathcal{E}_n\phi_n(x)$
($x=0,1,\ldots,x_{\text{max}}$) with
$\mathcal{H}=\hat{A}_{d_1}^{\dagger}\hat{A}_{d_1}+\tilde{\mathcal{E}}_{d_1}$
and \eqref{tHcPn=},
we have
\begin{equation*}
  \alpha B'(x)\nu(x+1)\check{P}_n(x+1)
  +\alpha D'(x)\nu(x-1)\check{P}_n(x-1)
  =\bigl(B(x)+D(x)-\mathcal{E}_n\bigr)\nu(x)\check{P}_n(x).
\end{equation*}
In terms of the functional relations of $\nu(x)$ \eqref{nurel}, it is
reduced to the original difference equation for $\check{P}_n(x)$ and it is
valid any $x\in\mathbb{C}$. 
By using this, we can show
\begin{equation*}
  \alpha B'(x)\text{W}[\check{\xi}_{d_1},\nu\check{P}_n](x)
  =\alpha D'(x)\text{W}[\check{\xi}_{d_1},\nu\check{P}_n](x-1)
  +(\tilde{\mathcal{E}}_{d_1}-\mathcal{E}_n)
  \check{\xi}_{d_1}(x)\nu(x)\check{P}_n(x).
%  \label{W[xi,n]s=0}
\end{equation*}
By setting $n=0$ and $x=0,1,\ldots,x_{\text{max}}$ in turn, we obtain
\begin{equation*}
  -\text{W}[\check{\xi}_{d_1},\nu](x)>0
  \ \ (x=0,1,\ldots,x_{\text{max}}).
\end{equation*}

Let us rewrite the deformed Hamiltonian $\mathcal{H}_{d_1}$ in the
standard form.
The potential functions $B_{d_1}(x)$ and $D_{d_1}(x)$ are introduced:
\begin{align}
  &B_{d_1}(x)\eqdef\alpha B'(x+1)
  \frac{\check{\xi}_{d_1}(x)}{\check{\xi}_{d_1}(x+1)}
  \frac{\text{W}[\check{\xi}_{d_1},\nu](x+1)}
  {\text{W}[\check{\xi}_{d_1},\nu](x)},\\
  &D_{d_1}(x)\eqdef\alpha D'(x)
  \frac{\check{\xi}_{d_1}(x+1)}{\check{\xi}_{d_1}(x)}
  \frac{\text{W}[\check{\xi}_{d_1},\nu](x-1)}
  {\text{W}[\check{\xi}_{d_1},\nu](x)}.
\end{align}
The positivity of $B_{d_1}(x)$ and $D_{d_1}(x)$ is shown above and
the boundary conditions $B_{d_1}(x_{\text{max}})=0$ and $D_{d_1}(0)=0$
are satisfied.
They satisfy the relations
\begin{align*}
  &B_{d_1}(x)D_{d_1}(x+1)
  =\hat{B}_{d_1}(x+1)\hat{D}_{d_1}(x+1),\\
  &B_{d_1}(x)+D_{d_1}(x)
  =\hat{B}_{d_1}(x)+\hat{D}_{d_1}(x+1)+\tilde{\mathcal{E}}_{d_1}.
\end{align*}
The standard form Hamiltonian is obtained:
\begin{align}
  &\mathcal{H}_{d_1}=\mathcal{A}_{d_1}^{\dagger}\mathcal{A}_{d_1},\\
  &\mathcal{A}_{d_1}\eqdef
  \sqrt{B_{d_1}(x)}-e^{\partial}\sqrt{D_{d_1}(x)},\quad
  \mathcal{A}_{d_1}^{\dagger}
  =\sqrt{B_{d_1}(x)}-\sqrt{D_{d_1}(x)}\,e^{-\partial},
\end{align}
in which $\mathcal{A}_{d_1}$ annihilates the groundstate eigenvector
\begin{equation}
  \mathcal{A}_{d_1}\phi_{d_10}(x)=0
  \ \ (x=0,1,\ldots,x_{\text{max}}).
\end{equation}

This one virtual state vector deletion is essentially the same procedure as
that developed for the exceptional orthogonal polynomials in \cite{os23}.
See \S\,\ref{sec:ef_miop_qR} for the explicit expressions.

\medskip

\noindent
\underline{multi virtual state vector deletion}

We repeat the above procedure and obtain the modified systems.
The number of deleted virtual state vectors should be less than or equal
$|\mathcal{V}|$ and $M$.

Let us assume that we have already deleted $s$ virtual state vectors
($s\geq 1$), which are labeled by $\{d_1,\ldots,d_s\}$
($d_j\in\mathcal{V}$ : mutually distinct).
Namely we have
\begin{align}
  &\mathcal{H}_{d_1\ldots d_s}\eqdef
  \hat{\mathcal{A}}_{d_1\ldots d_s}\hat{\mathcal{A}}_{d_1\ldots d_s}^{\dagger}
  +\tilde{\mathcal{E}}_{d_s},\quad
  \mathcal{H}_{d_1\ldots d_s}=(\mathcal{H}_{d_1\ldots d_s\,x,y})\quad
  (x,y=0,1,\ldots,x_{\text{max}}),
  \label{Hd1..ds}\\
  &\hat{\mathcal{A}}_{d_1\ldots d_s}\eqdef
  \sqrt{\hat{B}_{d_1\dots d_s}(x)}
  -e^{\partial}\sqrt{\hat{D}_{d_1\ldots d_s}(x)},
  \quad\hat{\mathcal{A}}_{d_1\ldots d_s}^{\dagger}=
  \sqrt{\hat{B}_{d_1\ldots d_s}(x)}
  -\sqrt{\hat{D}_{d_1\ldots d_s}(x)}\,e^{-\partial},\!\!\\
  &\hat{B}_{d_1\ldots d_s}(x)\eqdef\alpha B'(x+s-1)
  \frac{\text{W}[\check{\xi}_{d_1},\ldots,\check{\xi}_{d_{s-1}}](x)}
  {\text{W}[\check{\xi}_{d_1},\ldots,\check{\xi}_{d_{s-1}}](x+1)}\,
  \frac{\text{W}[\check{\xi}_{d_1},\ldots,\check{\xi}_{d_s}](x+1)}
  {\text{W}[\check{\xi}_{d_1},\ldots,\check{\xi}_{d_s}](x)},
  \label{Bdsform}\\
  &\hat{D}_{d_1\ldots d_s}(x)\eqdef\alpha D'(x)
  \frac{\text{W}[\check{\xi}_{d_1},\ldots,\check{\xi}_{d_{s-1}}](x+1)}
  {\text{W}[\check{\xi}_{d_1},\ldots,\check{\xi}_{d_{s-1}}](x)}\,
  \frac{\text{W}[\check{\xi}_{d_1},\ldots,\check{\xi}_{d_s}](x-1)}
  {\text{W}[\check{\xi}_{d_1},\ldots,\check{\xi}_{d_s}](x)},
  \label{Ddsform}\\
  &\phi_{d_1\ldots d_s\,n}(x)\eqdef
  \hat{\mathcal{A}}_{d_1\ldots d_s}\phi_{d_1\ldots d_{s-1}\,n}(x)
  \ \ (x=0,1,\ldots,x_{\text{max}};
  n=0,1,\ldots,n_{\text{max}}),\\
  &\tilde{\phi}_{d_1\ldots d_s\,\text{v}}(x)\eqdef
  \hat{\mathcal{A}}_{d_1\ldots d_s}
  \tilde{\phi}_{d_1\ldots d_{s-1}\,\text{v}}(x)
  +\delta_{x,x_{\text{max}}}\varphi_{d_1\ldots d_s\,\text{v}},
  \ (x=0,1,\ldots,x_{\text{max}}; 
  \text{v}\in\mathcal{V}\backslash\{d_1,\ldots,d_s\}),\n
  &\qquad
  \varphi_{d_1\ldots d_s\,\text{v}}\eqdef\phi_{d_1\ldots d_s\,0}(x_{\text{max}})
  \frac{\text{W}[\check{\xi}_{d_1},\ldots,\check{\xi}_{d_{s}}](x_{\text{max}})
  \text{W}[\check{\xi}_{d_1},\ldots,\check{\xi}_{d_{s-1}},
  \check{\xi}_{\text{v}}](x_{\text{max}}+1)}
  {\text{W}[\check{\xi}_{d_1},\ldots,\check{\xi}_{d_{s-1}}](x_{\text{max}}+1)
  \text{W}[\check{\xi}_{d_1},\ldots,\check{\xi}_{d_s},\nu](x_{\text{max}})},\\
  &\mathcal{H}_{d_1\ldots d_s}\phi_{d_1\ldots d_s\,n}(x)
  =\mathcal{E}_n\phi_{d_1\ldots d_s\,n}(x)
  \ \ (x=0,1,\ldots,x_{\text{max}};
  n=0,1,\ldots,n_{\text{max}}),
  \label{Hd1..dsphid1..ds=}\\
  &\mathcal{H}_{d_1\ldots d_s}\tilde{\phi}_{d_1\ldots d_s\,\text{v}}(x)
  =\tilde{\mathcal{E}}_\text{v}\tilde{\phi}_{d_1\ldots d_s\,\text{v}}(x)
  \ \ (x=0,1,\ldots,x_{\text{max}}-1;
  \text{v}\in\mathcal{V}\backslash\{d_1,\ldots,d_s\}),
  \label{Hd1..dstphid1..ds=}\\
  &(\phi_{d_1\ldots d_s\,n},\phi_{d_1\ldots d_s\,m})
  \eqdef\sum_{x=0}^{x_{\text{max}}}
  \phi_{d_1\ldots d_s\,n}(x)\phi_{d_1\ldots d_s\,m}(x)
  =\prod_{j=1}^s(\mathcal{E}_n-\tilde{\mathcal{E}}_{d_j})\cdot
  \frac{1}{d_n^2}\delta_{nm}\n
  &\qquad\qquad\qquad\qquad\qquad\qquad\qquad\qquad\qquad\qquad\qquad
  (n,m=0,1,\ldots,n_{\text{max}}).
  \label{(phid1..dsm,phid1..dsn)}
\end{align}
The eigenvectors and the virtual state vectors have Casoratian expressions
($x=0,1,\ldots,x_{\text{max}}$):
\begin{align} 
  &\phi_{d_1\ldots d_s\,n}(x)=
  \frac{(-1)^s\sqrt{\prod_{j=1}^s\alpha B'(x+j-1)}
  \,\tilde{\phi}_0(x)\,
  \text{W}[\check{\xi}_{d_1},\ldots,\check{\xi}_{d_s},\nu\check{P}_n](x)}
  {\sqrt{\text{W}[\check{\xi}_{d_1},\ldots,\check{\xi}_{d_s}](x)\,
  \text{W}[\check{\xi}_{d_1},\ldots,\check{\xi}_{d_s}](x+1)}},
  \label{phid1..dsn}\\[2pt]
  &\tilde{\phi}_{d_1\ldots d_s\,\text{v}}(x)=
  \frac{(-1)^s\sqrt{\prod_{j=1}^s\alpha B'(x+j-1)}
  \,\tilde{\phi}_0(x)\,
  \text{W}[\check{\xi}_{d_1},\ldots,\check{\xi}_{d_s},
  \check{\xi}_{\text{v}}](x)}
  {\sqrt{\text{W}[\check{\xi}_{d_1},\ldots,\check{\xi}_{d_s}](x)\,
  \text{W}[\check{\xi}_{d_1},\ldots,\check{\xi}_{d_s}](x+1)}}.
  \label{phitd1..dsv}
\end{align}
The Casoratian in the virtual state vectors
$\text{W}[\check{\xi}_{d_1},\ldots,\check{\xi}_{d_s},
\check{\xi}_{\text{v}}](x)$ has definite sign for
$x=0,1,\ldots,x_{\text{max}}+1$, and that appearing in the groundstate
eigenvector $\text{W}[\check{\xi}_{d_1},\ldots,\check{\xi}_{d_s},\nu](x)$
has definite sign for $x=0,1,\ldots,x_{\text{max}}$, too.

The next step begins with rewriting the Hamiltonian
$\mathcal{H}_{d_1\ldots d_s}$ by choosing the next virtual state to be
deleted $d_{s+1}\in\mathcal{V}\backslash\{d_1,\ldots,d_s\}$.
The potential functions
$\hat{B}_{d_1\ldots d_{s+1}}(x)$ and $\hat{D}_{d_1\ldots d_{s+1}}(x)$
are defined as in \eqref{Bdsform}--\eqref{Ddsform} by  $s\to s+1$.
We have $\hat{B}_{d_1\ldots d_{s+1}}(x)>0$
($x=0,1,\ldots,x_{\text{max}}$),
$\hat{D}_{d_1\ldots d_{s+1}}(0)=
\hat{D}_{d_1\ldots d_{s+1}}(x_{\text{max}}+1)=0$,
$\hat{D}_{d_1\ldots d_{s+1}}(x)>0$ ($x=1,2,\ldots,x_{\text{max}}$).
These functions satisfy the relations
\begin{align*}
  &\hat{B}_{d_1\ldots d_{s+1}}(x)\hat{D}_{d_1\ldots d_{s+1}}(x+1)
  =\hat{B}_{d_1\ldots d_s}(x+1)\hat{D}_{d_1\ldots d_s}(x+1),\\
  &\hat{B}_{d_1\ldots d_{s+1}}(x)+\hat{D}_{d_1\ldots d_{s+1}}(x)
  +\tilde{\mathcal{E}}_{d_{s+1}}
  =\hat{B}_{d_1\ldots d_s}(x)+\hat{D}_{d_1\ldots d_s}(x+1)
  +\tilde{\mathcal{E}}_{d_s}.
\end{align*}
The Hamiltonian $\mathcal{H}_{d_1\ldots d_s}$ is rewritten as:
\begin{gather*}
  \mathcal{H}_{d_1\ldots d_s}
  =\hat{\mathcal{A}}_{d_1\ldots d_{s+1}}^{\dagger}
  \hat{\mathcal{A}}_{d_1\ldots d_{s+1}}
  +\tilde{\mathcal{E}}_{d_{s+1}},\\
  \hat{\mathcal{A}}_{d_1\ldots d_{s+1}}\!\eqdef\!
  \sqrt{\hat{B}_{d_1\ldots d_{s+1}}(x)}
  -e^{\partial}\sqrt{\hat{D}_{d_1\ldots d_{s+1}}(x)},\
  \hat{\mathcal{A}}_{d_1\ldots d_{s+1}}^{\dagger}\!
  =\!\sqrt{\hat{B}_{d_1\ldots d_{s+1}}(x)}
  -\sqrt{\hat{D}_{d_1\ldots d_{s+1}}(x)}\,e^{-\partial}.
\end{gather*}

Now let us define a new Hamiltonian $\mathcal{H}_{d_1\ldots d_{s+1}}$
by changing the orders of
$\hat{\mathcal{A}}_{d_1\ldots d_{s+1}}^{\dagger}$ and
$\hat{\mathcal{A}}_{d_1\ldots d_{s+1}}$ together with 
the eigenvectors $\phi_{d_1\ldots d_{s+1}\,n}(x)$ and
the virtual state vectors $\tilde{\phi}_{d_1\ldots d_{s+1}\,\text{v}}(x)$:
\begin{align*}
  &\mathcal{H}_{d_1\ldots d_{s+1}}\eqdef
  \hat{\mathcal{A}}_{d_1\ldots d_{s+1}}
  \hat{\mathcal{A}}_{d_1\ldots d_{s+1}}^{\dagger}
  +\tilde{\mathcal{E}}_{d_{s+1}},\quad
  \mathcal{H}_{d_1\ldots d_{s+1}}=(\mathcal{H}_{d_1\ldots d_{s+1}\,x,y})
  \ \ (x,y=0,1,\ldots,x_{\text{max}}),\\
  &\phi_{d_1\ldots d_{s+1}\,n}(x)\eqdef
  \hat{\mathcal{A}}_{d_1\ldots d_{s+1}}\phi_{d_1\ldots d_s\,n}(x)
  \ \ (x=0,1,\ldots,x_{\text{max}}; n=0,1,\ldots,n_{\text{max}}),\\
  &\tilde{\phi}_{d_1\ldots d_{s+1}\,\text{v}}(x)\eqdef
  \hat{\mathcal{A}}_{d_1\ldots d_{s+1}}
  \tilde{\phi}_{d_1\ldots d_s\,\text{v}}(x)
  +\delta_{x,x_{\text{max}}}\varphi_{d_1\ldots d_{s+1}\,\text{v}},\\
  &\hspace{40mm} \ (x=0,1,\ldots,x_{\text{max}};
  \text{v}\in\mathcal{V}\backslash\{d_1,\ldots,d_{s+1}\}),\\
  &\qquad
  \varphi_{d_1\ldots d_{s+1}\,\text{v}}\eqdef
  \phi_{d_1\ldots d_{s+1}\,0}(x_{\text{max}})
  \frac{\text{W}[\check{\xi}_{d_1},\ldots,\check{\xi}_{d_{s+1}}](x_{\text{max}})
  \text{W}[\check{\xi}_{d_1},\ldots,\check{\xi}_{d_{s}},
  \check{\xi}_{\text{v}}](x_{\text{max}}+1)}
  {\text{W}[\check{\xi}_{d_1},\ldots,\check{\xi}_{d_{{s}}}](x_{\text{max}}+1)
  \text{W}[\check{\xi}_{d_1},\ldots,\check{\xi}_{d_{s+1}},\nu](x_{\text{max}})}.
\end{align*}
The orthogonality relation reads
\begin{equation*}
  \quad(\phi_{d_1\ldots d_{s+1}\,n},\phi_{d_1\ldots d_{s+1}\,m})
  =\prod_{j=1}^{s+1}(\mathcal{E}_n-\tilde{\mathcal{E}}_{d_j})
  \cdot\frac{1}{d_n^2}\delta_{nm}
  \ \ (n,m=0,1,\ldots,n_{\text{max}}).
\end{equation*}

The functions $\phi_{d_1\ldots d_{s+1}\,n}(x)$
and $\tilde{\phi}_{d_1\ldots d_{s+1}\,\text{v}}(x)$ are expressed as
Casoratians as in \eqref{phid1..dsn}--\eqref{phitd1..dsv}.
The Casoratian $\text{W}[\check{\xi}_{d_1},\ldots,\check{\xi}_{d_{s+1}},
\check{\xi}_{\text{v}}](x)$ has definite sign 
\begin{align*}
  &\pm\frac{\text{W}[\check{\xi}_{d_1},\ldots,\check{\xi}_{d_s}](0)
  (\tilde{\mathcal{E}}_{d_{s+1}}-\tilde{\mathcal{E}}_{\text{v}})}
  {\text{W}[\check{\xi}_{d_1},\ldots,\check{\xi}_{d_{s+1}}](0)
  \text{W}[\check{\xi}_{d_1},\ldots,\check{\xi}_{d_s},
  \check{\xi}_{\text{v}}](0)}>0\n
  &\Rightarrow
  \pm\text{W}[\check{\xi}_{d_1},\ldots,\check{\xi}_{d_{s+1}},
  \check{\xi}_{\text{v}}](x)>0
  \ \ (x=0,1,\ldots,x_{\text{max}}+1).
\end{align*}
Likewise $\text{W}[\check{\xi}_{d_1},\ldots,\check{\xi}_{d_{s+1}},\nu](x)$
and the lowest eigenvector $\phi_{d_1\ldots d_{s+1}\,0}(x)$ have definite
sign
\begin{align*}
  &\mp\frac{\text{W}[\check{\xi}_{d_1},\ldots,\check{\xi}_{d_s}](0)}
  {\text{W}[\check{\xi}_{d_1},\ldots,\check{\xi}_{d_{s+1}}](0)
  \text{W}[\check{\xi}_{d_1},\ldots,\check{\xi}_{d_s},\nu](0)}>0\n
  &\Rightarrow
  \pm\text{W}[\check{\xi}_{d_1},\ldots,\check{\xi}_{d_{s+1}},\nu](x)>0
  \ \ (x=0,1,\ldots,x_{\text{max}}).
\end{align*}
These establish the $s+1$ case.

At the end of this subsection we present this deformed Hamiltonian
$\mathcal{H}_{d_1\ldots d_s}$ in the standard form, in which the
$\mathcal{A}$ operator annihilates the groundstate eigenvector:
\begin{align}
  &\mathcal{H}_{d_1\ldots d_s}
  =\mathcal{A}_{d_1\ldots d_s}^{\dagger}\mathcal{A}_{d_1\ldots d_s},
  \label{sstanham1}\\
  &\mathcal{A}_{d_1\ldots d_s}\eqdef
  \sqrt{B_{d_1\ldots d_s}(x)}
  -e^{\partial}\sqrt{D_{d_1\ldots d_s}(x)},
  \ \ \mathcal{A}_{d_1\ldots d_s}^{\dagger}
  =\sqrt{B_{d_1\ldots d_s}(x)}
  -\sqrt{D_{d_1\ldots d_s}(x)}\,e^{-\partial},
  \label{sstanham2}
\end{align}
which satisfies
\begin{equation}
  \mathcal{A}_{d_1\ldots d_s}\phi_{d_1\ldots d_s\,0}(x)=0
  \ \ (x=0,1,\ldots,x_{\text{max}}).
\end{equation}
The potential functions $B_{d_1\ldots d_s}(x)$ and
$D_{d_1\ldots d_s}(x)$ are:
\begin{align}
  &B_{d_1\ldots d_s}(x)\eqdef\alpha B'(x+s)
  \frac{\text{W}[\check{\xi}_{d_1},\ldots,\check{\xi}_{d_s}](x)}
  {\text{W}[\check{\xi}_{d_1},\ldots,\check{\xi}_{d_s}](x+1)}\,
  \frac{\text{W}[\check{\xi}_{d_1},\ldots,\check{\xi}_{d_s},\nu](x+1)}
  {\text{W}[\check{\xi}_{d_1},\ldots,\check{\xi}_{d_s},\nu](x)},
  \label{Bd1..ds}\\
  &D_{d_1\ldots d_s}(x)\eqdef\alpha D'(x)
  \frac{\text{W}[\check{\xi}_{d_1},\ldots,\check{\xi}_{d_s}](x+1)}
  {\text{W}[\check{\xi}_{d_1},\ldots,\check{\xi}_{d_s}](x)}\,
  \frac{\text{W}[\check{\xi}_{d_1},\ldots,\check{\xi}_{d_s},\nu](x-1)}
  {\text{W}[\check{\xi}_{d_1},\ldots,\check{\xi}_{d_s},\nu](x)}.
  \label{Dd1..ds}
\end{align}
The positivity of $B_{d_1\ldots d_s}(x)$ and $D_{d_1\ldots d_s}(x)$
is shown above and the boundary conditions
$B_{d_1\ldots d_s}(x_{\text{max}})=0$ and $D_{d_1\ldots d_s}(0)=0$ are
satisfied. They satisfy the relations
\begin{align*}
  &B_{d_1\ldots d_s}(x)D_{d_1\ldots d_s}(x+1)
  =\hat{B}_{d_1\ldots d_s}(x+1)\hat{D}_{d_1\ldots d_s}(x+1),
%  \label{ss1iden1}
  \\
  &B_{d_1\ldots d_s}(x)+D_{d_1\ldots d_s}(x)
  =\hat{B}_{d_1\ldots d_s}(x)+\hat{D}_{d_1\ldots d_s}(x+1)
  +\tilde{\mathcal{E}}_{d_s}.
%  \label{ss1iden2}
\end{align*}

It should be stressed that the above results after $s$-deletions are
independent of the orders of deletions ($\phi_{d_1\ldots d_s\,n}(x)$
and $\tilde{\phi}_{d_1\ldots d_s\,\text{v}}(x)$ may change sign).

%%%%%%%%%%%%%%%%%%%%%%%%%%%%%%%%%%%%%%%%%%%%%%%%%%%%%%%%%%%%%%%%%%
%                                                                %
% 3.2 Explicit forms of multi-indexed ($q$-)Racah polynomials    %
%                                                                %
%%%%%%%%%%%%%%%%%%%%%%%%%%%%%%%%%%%%%%%%%%%%%%%%%%%%%%%%%%%%%%%%%%
\subsection{Explicit forms of multi-indexed ($q$-)Racah polynomials}
\label{sec:ef_miop_qR}

Here we present the main results of the paper.
The multi-indexed ($q$-)Racah  polynomials are obtained by applying
the method of virtual states deletion to the ($q$-)Racah system.
The parameter $\bm{\lambda}=(\lambda_1,\lambda_2,\ldots)$ dependence is
now shown explicitly.
The eigenvectors of the models in \S\,5 of \cite{os12} are described by
orthogonal polynomials in the sinusoidal coordinate $\eta(x;\bm{\lambda})$.
The auxiliary function $\varphi(x;\bm{\lambda})$ is defined by
\begin{equation}
  \varphi(x;\bm{\lambda})\eqdef
  \frac{\eta(x+1;\bm{\lambda})-\eta(x;\bm{\lambda})}{\eta(1;\bm{\lambda})},
  \label{varphidef}
\end{equation}
and it satisfies (with $\bm{\delta}$ defined in
\eqref{lamdelR}--\eqref{lamdelqR})
\begin{equation}
  \frac{\varphi(x;\bm{\lambda})}{\varphi(x-1;\bm{\lambda}+2\bm{\delta})}
  =\varphi(1;\bm{\lambda}).
  \label{varphiid}
\end{equation}
All the models in \S\,5 of \cite{os12} have shape invariance \cite{genden}.
The following relations are very useful:
\begin{align}
  &\varphi(x;\bm{\lambda})=\sqrt{\frac{B(0;\bm{\lambda})}{B(x;\bm{\lambda})}}
  \,\frac{\phi_0(x;\bm{\lambda}+\bm{\delta})}{\phi_0(x;\bm{\lambda})},\quad
  \varphi(x;\bm{\lambda})=\sqrt{\frac{B(0;\bm{\lambda})}{D(x+1;\bm{\lambda})}}
  \,\frac{\phi_0(x;\bm{\lambda}+\bm{\delta})}{\phi_0(x+1;\bm{\lambda})},
  \label{OS12(4.12,13)}\\
  &\frac{B(x;\bm{\lambda}+\bm{\delta})}{B(x+1;\bm{\lambda})}
  =\kappa^{-1}\frac{\varphi(x+1;\bm{\lambda})}{\varphi(x;\bm{\lambda})},\quad
  \frac{D(x;\bm{\lambda}+\bm{\delta})}{D(x;\bm{\lambda})}
  =\kappa^{-1}\frac{\varphi(x-1;\bm{\lambda})}{\varphi(x;\bm{\lambda})}.
  \label{OS12(5.81etc)}
\end{align}

We delete $M$ virtual state vectors labeled by
\begin{equation}
  \mathcal{D}=\{d_1,d_2,\ldots,d_M\}
  \ \ (d_j\in\mathcal{V} : \text{mutually distinct}),
\end{equation}
and denote $\mathcal{H}_{d_1\ldots d_M}$, $\phi_{d_1\ldots d_M\,n}$,
$\mathcal{A}_{d_1\ldots d_M}$, etc. by  $\mathcal{H}_{\mathcal{D}}$,
$\phi_{\mathcal{D}\,n}$, $\mathcal{A}_{\mathcal{D}}$, etc.

Let us denote the eigenvector $\phi_{\mathcal{D}\,n}(x;\bm{\lambda})$
in \eqref{phid1..dsn} after $M$ deletions ($s=M$) by
$\phi^{\text{gen}}_{\mathcal{D}\,n}(x;\bm{\lambda})$.
We define two polynomials $\check{\Xi}_{\mathcal{D}}(x;\bm{\lambda})$ and
$\check{P}_{\mathcal{D},n}(x;\bm{\lambda})$, to be called the denominator
polynomial and the multi-indexed orthogonal polynomial, respectively,
from the Casoratians as follows:
\begin{align}
  &\text{W}[\check{\xi}_{d_1},\ldots,\check{\xi}_{d_M}](x;\bm{\lambda})
  =\mathcal{C}_{\mathcal{D}}(\bm{\lambda})\varphi_M(x;\bm{\lambda})
  \check{\Xi}_{\mathcal{D}}(x;\bm{\lambda}),
  \label{XiDdef}\\
  &\text{W}[\check{\xi}_{d_1},\ldots,\check{\xi}_{d_M},\nu\check{P}_n]
  (x;\bm{\lambda})
  =\mathcal{C}_{\mathcal{D},n}(\bm{\lambda})
  \varphi_{M+1}(x;\bm{\lambda})
  \check{P}_{\mathcal{D},n}(x;\bm{\lambda})
  \nu(x;\bm{\lambda}+M\tilde{\bm{\delta}}),
  \label{cPDndef}\\
  &\tilde{\bm{\delta}}\eqdef(0,0,1,1),\ \quad
  \mathfrak{t}(\bm{\lambda})+\beta\bm{\delta}=
  \mathfrak{t}(\bm{\lambda}+\beta\tilde{\bm{\delta}})
  \ \ (\forall\beta\in\mathbb{R}).
  \label{deltat}
\end{align}
The constants $\mathcal{C}_{\mathcal{D}}(\bm{\lambda})$ and
$\mathcal{C}_{\mathcal{D},n}(\bm{\lambda})$ are specified later.
The auxiliary function $\varphi_M(x;\bm{\lambda})$ is defined by \cite{os22}:
\begin{align}
  \varphi_M(x;\bm{\lambda})&\eqdef\prod_{1\leq j<k\leq M}
  \frac{\eta(x+k-1;\bm{\lambda})-\eta(x+j-1;\bm{\lambda})}
  {\eta(k-j;\bm{\lambda})}\n
  &=\prod_{1\leq j<k\leq M}
  \varphi\bigl(x+j-1;\bm{\lambda}+(k-j-1)\bm{\delta}\bigr),
  \label{varphiMdef}
\end{align}
and $\varphi_0(x;\bm{\lambda})=\varphi_1(x;\bm{\lambda})=1$.
The eigenvector \eqref{phid1..dsn} is rewritten as
\begin{align}
  \phi^{\text{gen}}_{\mathcal{D}\,n}(x;\bm{\lambda})
  &=(-1)^M\kappa^{\frac14M(M-1)}
  \frac{\mathcal{C}_{\mathcal{D},n}(\bm{\lambda})}
  {\mathcal{C}_{\mathcal{D}}(\bm{\lambda})}
  \sqrt{\prod_{j=1}^M\alpha(\bm{\lambda})
  B'\bigl(0;\bm{\lambda}+(j-1)\tilde{\bm{\delta}}\bigr)}\n
  &\quad\times
  \frac{\phi_0(x;\bm{\lambda}+M\tilde{\bm{\delta}})}
  {\sqrt{\check{\Xi}_{\mathcal{D}}(x;\bm{\lambda})
  \check{\Xi}_{\mathcal{D}}(x+1;\bm{\lambda})}}
  \check{P}_{\mathcal{D},n}(x;\bm{\lambda}).
\end{align}
The multi-indexed orthogonal polynomial
$\check{P}_{\mathcal{D},n}(x;\bm{\lambda})$ \eqref{cPDndef} has an expression
\begin{align}
  \check{P}_{\mathcal{D},n}(x;\bm{\lambda})
  &=\mathcal{C}_{\mathcal{D},n}(\bm{\lambda})^{-1}
  \varphi_{M+1}(x;\bm{\lambda})^{-1}\n
  &\quad\times\left|
  \begin{array}{cccc}
  \check{\xi}_{d_1}(x_1)&\cdots&\check{\xi}_{d_M}(x_1)
  &r_1(x_1)\check{P}_n(x_1)\\
  \check{\xi}_{d_1}(x_2)&\cdots&\check{\xi}_{d_M}(x_2)
  &r_2(x_2)\check{P}_n(x_2)\\
  \vdots&\cdots&\vdots&\vdots\\
  \check{\xi}_{d_1}(x_{M+1})&\cdots&\check{\xi}_{d_M}(x_{M+1})
  &r_{M+1}(x_{M+1})\check{P}_n(x_{M+1})\\
  \end{array}\right|,
  \label{cPDn}
\end{align}
where $x_j\eqdef x+j-1$ and $r_j(x)=r_j(x;\bm{\lambda},M)$
($1\leq j\leq M+1$) are given by
\begin{equation}
  r_j\bigl(x+j-1;\bm{\lambda},M\bigr)
  \eqdef\left\{
  \begin{array}{ll}
  {\displaystyle
  \frac{(x+a,x+b)_{j-1}(x+d-a+j,x+d-b+j)_{M+1-j}}
  {(d-a+1,d-b+1)_M}}&:\text{R}\\[8pt]
  {\displaystyle
  \frac{(aq^x,bq^x;q)_{j-1}(a^{-1}dq^{x+j},b^{-1}dq^{x+j};q)_{M+1-j}}
  {(abd^{-1}q^{-1})^{j-1}q^{Mx}(a^{-1}dq,b^{-1}dq;q)_M}}&:\text{$q$R}
  \end{array}\right..
  \label{rj}
\end{equation}
One can show that $\check{\Xi}_{\mathcal{D}}$ \eqref{XiDdef} and
$\check{P}_{\mathcal{D},n}$ \eqref{cPDn} are indeed polynomials in $\eta$:
\begin{equation}
  \check{\Xi}_{\mathcal{D}}(x;\bm{\lambda})\eqdef
  \Xi_{\mathcal{D}}\bigl(\eta(x;\bm{\lambda}+(M-1)\tilde{\bm{\delta}});
  \bm{\lambda}\bigr),
  \quad
  \check{P}_{\mathcal{D},n}(x;\bm{\lambda})\eqdef
  P_{\mathcal{D},n}\bigl(\eta(x;\bm{\lambda}+M\tilde{\bm{\delta}});
  \bm{\lambda}\bigr),
  \label{XiP_poly}
\end{equation}
and their degrees are generically $\ell$ and $\ell+n$, respectively
(see \eqref{c_PDn}).
Here $\ell$ is
\begin{equation}
  \ell\eqdef\sum_{j=1}^Md_j-\tfrac12M(M-1).
  \label{lform}
\end{equation}
The involution properties \eqref{remark} of these polynomials are the
consequence of those of the basic polynomials $\check{P}_n(x)$ and
$\check{\xi}_{d_j}(x)$.
We adopt the standard normalisation for $\check{\Xi}_{\mathcal{D}}$ and
$\check{P}_{\mathcal{D},n}$:
$\check{\Xi}_{\mathcal{D}}(0;\bm{\lambda})=1$,
$\check{P}_{\mathcal{D},n}(0;\bm{\lambda})=1$,
which determine the constants $\mathcal{C}_{\mathcal{D}}(\bm{\lambda})$
and $\mathcal{C}_{\mathcal{D},n}(\bm{\lambda})$,
\begin{align}
  \mathcal{C}_{\mathcal{D}}(\bm{\lambda})&\eqdef
  \frac{1}{\varphi_M(0;\bm{\lambda})}
  \prod_{1\leq j<k\leq M}
  \frac{\tilde{\mathcal{E}}_{d_j}(\bm{\lambda})
  -\tilde{\mathcal{E}}_{d_k}(\bm{\lambda})}
  {\alpha(\bm{\lambda})B'(j-1;\bm{\lambda})},
  \label{CD}\\
  \mathcal{C}_{\mathcal{D},n}(\bm{\lambda})&\eqdef
  (-1)^M\mathcal{C}_{\mathcal{D}}(\bm{\lambda})
  \tilde{d}_{\mathcal{D},n}(\bm{\lambda})^2,\quad
  \tilde{d}_{\mathcal{D},n}(\bm{\lambda})^2\eqdef
  \frac{\varphi_M(0;\bm{\lambda})}{\varphi_{M+1}(0;\bm{\lambda})}
  \prod_{j=1}^M\frac{\mathcal{E}_n(\bm{\lambda})
  -\tilde{\mathcal{E}}_{d_j}(\bm{\lambda})}
  {\alpha(\bm{\lambda})B'(j-1;\bm{\lambda})}.
  \label{CDn}
\end{align}
The use of {\em dual polynomials\/} $Q_x(\mathcal{E}_n)
\eqdef P_n(\eta(x))$ \cite{os12} is essential for the
derivation of these results. The three term recurrence relations of
$\{Q_x(\mathcal{E})\}$ are specified by $B(x)$ and $D(x)$.
The denominator polynomial $\check{\Xi}_{\mathcal{D}}(x;\bm{\lambda})$
is positive for $x=0,1,\ldots,x_{\text{max}}+1$.
The lowest degree multi-indexed orthogonal polynomial
$\check{P}_{\mathcal{D},0}(x;\bm{\lambda})$ is related to
$\check{\Xi}_{\mathcal{D}}(x;\bm{\lambda})$ by the parameter shift
$\bm{\lambda}\to\bm{\lambda}+\bm{\delta}$:
\begin{equation}
  \check{P}_{\mathcal{D},0}(x;\bm{\lambda})
  =\check{\Xi}_{\mathcal{D}}(x;\bm{\lambda}+\bm{\delta}).
  \label{PD0=XiD}
\end{equation}
The potential functions $B_{\mathcal{D}}$ and $D_{\mathcal{D}}$
\eqref{Bd1..ds}--\eqref{Dd1..ds} after $M$-deletion ($s=M$) can be
expressed neatly in terms of the denominator polynomial:
\begin{align}
  B_{\mathcal{D}}(x;\bm{\lambda})&=B(x;\bm{\lambda}+M\tilde{\bm{\delta}})\,
  \frac{\check{\Xi}_{\mathcal{D}}(x;\bm{\lambda})}
  {\check{\Xi}_{\mathcal{D}}(x+1;\bm{\lambda})}
  \frac{\check{\Xi}_{\mathcal{D}}(x+1;\bm{\lambda}+\bm{\delta})}
  {\check{\Xi}_{\mathcal{D}}(x;\bm{\lambda}+\bm{\delta})},
  \label{BD2}\\
  D_{\mathcal{D}}(x;\bm{\lambda})&=D(x;\bm{\lambda}+M\tilde{\bm{\delta}})\,
  \frac{\check{\Xi}_{\mathcal{D}}(x+1;\bm{\lambda})}
  {\check{\Xi}_{\mathcal{D}}(x;\bm{\lambda})}
  \frac{\check{\Xi}_{\mathcal{D}}(x-1;\bm{\lambda}+\bm{\delta})}
  {\check{\Xi}_{\mathcal{D}}(x;\bm{\lambda}+\bm{\delta})}.
  \label{DD2}
\end{align}
These formulas look similar to those in the exceptional polynomials
\cite{os23}.
The groundstate eigenvector $\phi_{\mathcal{D}\,0}$ is expressed by
$\phi_0(x)$ \eqref{phi0=prodB/D} and
$\check{\Xi}_{\mathcal{D}}(x;\bm{\lambda})$:
\begin{align}
  \phi_{\mathcal{D}\,0}(x;\bm{\lambda})&=
  \sqrt{\prod_{y=0}^{x-1}\frac{B_{\mathcal{D}}(y)}{D_{\mathcal{D}}(y+1)}}
  =\phi_0(x;\bm{\lambda}+M\tilde{\bm{\delta}})
  \sqrt{\frac{\check{\Xi}_{\mathcal{D}}(1;\bm{\lambda})}
  {\check{\Xi}_{\mathcal{D}}(x;\bm{\lambda})
  \check{\Xi}_{\mathcal{D}}(x+1;\bm{\lambda})}}\,
  \check{\Xi}_{\mathcal{D}}(x;\bm{\lambda}+\bm{\delta})\n
  &=\psi_{\mathcal{D}}(x;\bm{\lambda})
  \check{P}_{\mathcal{D},0}(x;\bm{\lambda})
  \propto\phi^{\text{gen}}_{\mathcal{D}\,0}(x;\bm{\lambda}),\\
  \psi_{\mathcal{D}}(x;\bm{\lambda})&\eqdef
  \sqrt{\check{\Xi}_{\mathcal{D}}(1;\bm{\lambda})}\,
  \frac{\phi_0(x;\bm{\lambda}+M\tilde{\bm{\delta}})}
  {\sqrt{\check{\Xi}_{\mathcal{D}}(x;\bm{\lambda})\,
  \check{\Xi}_{\mathcal{D}}(x+1;\bm{\lambda})}},\quad
  \psi_{\mathcal{D}}(0;\bm{\lambda})=1.
\end{align}
We arrive at the normalised eigenvector
$\phi_{\mathcal{D}\,n}(x;\bm{\lambda})$ with the orthogonality relation,
\begin{align}
  &\phi_{\mathcal{D}\,n}(x;\bm{\lambda})
  \eqdef\psi_{\mathcal{D}}(x;\bm{\lambda})
  \check{P}_{\mathcal{D},n}(x;\bm{\lambda})
  \propto\phi^{\text{gen}}_{\mathcal{D}\,n}(x;\bm{\lambda}),\quad
  \phi_{\mathcal{D}\,n}(0;\bm{\lambda})=1,\\
  &\sum_{x=0}^{x_{\text{max}}}
%  \frac{\phi_0(x;\bm{\lambda}+M\tilde{\bm{\delta}})^2}
%  {\check{\Xi}_{\mathcal{D}}(x;\bm{\lambda})
%  \check{\Xi}_{\mathcal{D}}(x+1;\bm{\lambda})}\,
  \frac{\psi_{\mathcal{D}}(x;\bm{\lambda})^2}
  {\check{\Xi}_{\mathcal{D}}(1;\bm{\lambda})}
  \check{P}_{\mathcal{D},n}(x;\bm{\lambda})
  \check{P}_{\mathcal{D},m}(x;\bm{\lambda})
  =\frac{\delta_{nm}}{d_n(\bm{\lambda})^2
  \tilde{d}_{\mathcal{D},n}(\bm{\lambda})^2}
  \ \ (n,m=0,1,\ldots,n_{\text{max}}).
\end{align}
It is worthwhile to emphasise that the above orthogonality relation is
a rational equation of $\bm{\lambda}$ or $q^{\bm{\lambda}}$, and it is
valid for any value of $\bm{\lambda}$ (except for the zeros of denominators)
but the weight function may not be positive definite.

The shape invariance of the original system is inherited by
the deformed systems.
The matrix $\hat{\mathcal{A}}_{d_1\ldots d_{s+1}}(\bm{\lambda})$ intertwines
the two Hamiltonians $\mathcal{H}_{d_1\ldots d_s}(\bm{\lambda})$ and
$\mathcal{H}_{d_1\ldots d_{s+1}}(\bm{\lambda})$,
\begin{align*}
  \hat{\mathcal{A}}_{d_1\ldots d_{s+1}}(\bm{\lambda})^{\dagger}
  \hat{\mathcal{A}}_{d_1\ldots d_{s+1}}(\bm{\lambda})
  &=\mathcal{H}_{d_1\ldots d_s}(\bm{\lambda})
  -\tilde{\mathcal{E}}_{d_{s+1}}(\bm{\lambda}),\\
  \hat{\mathcal{A}}_{d_1\ldots d_{s+1}}(\bm{\lambda})
  \hat{\mathcal{A}}_{d_1\ldots d_{s+1}}(\bm{\lambda})^{\dagger}
  &=\mathcal{H}_{d_1\ldots d_{s+1}}(\bm{\lambda})
  -\tilde{\mathcal{E}}_{d_{s+1}}(\bm{\lambda}),
\end{align*}
and it has the inverse.
By the same argument given in \S\,4 of \cite{os20}, the shape invariance of
$\mathcal{H}(\bm{\lambda})$ is inherited by $\mathcal{H}_{d_1}(\bm{\lambda})$,
$\mathcal{H}_{d_1d_2}(\bm{\lambda})$, $\cdots$.
Therefore the Hamiltonian $\mathcal{H}_{\mathcal{D}}(\bm{\lambda})$
is shape invariant:
\begin{equation}
  \mathcal{A}_{\mathcal{D}}(\bm{\lambda})
  \mathcal{A}_{\mathcal{D}}(\bm{\lambda})^{\dagger}
  =\kappa\mathcal{A}_{\mathcal{D}}(\bm{\lambda}+\bm{\delta})^{\dagger}
  \mathcal{A}_{\mathcal{D}}(\bm{\lambda}+\bm{\delta})
  +\mathcal{E}_1(\bm{\lambda}).
  \label{shapeinvD}
\end{equation}
As a consequence of the shape invariance and the normalisation,
the actions of $\mathcal{A}_{\mathcal{D}}(\bm{\lambda})$ and
$\mathcal{A}_{\mathcal{D}}(\bm{\lambda})^{\dagger}$ on the eigenvectors
$\phi_{\mathcal{D}\,n}(x;\bm{\lambda})$ are
\begin{align}
  &\mathcal{A}_{\mathcal{D}}(\bm{\lambda})
  \phi_{\mathcal{D}\,n}(x;\bm{\lambda})
  =\frac{\mathcal{E}_n(\bm{\lambda})}
  {\sqrt{B_{\mathcal{D}}(0;\bm{\lambda})}}\,
  \phi_{\mathcal{D}\,n-1}(x;\bm{\lambda}+\bm{\delta})
  \ \ (x=0,1,\ldots,x_{\text{max}}-1),
  \label{ADphiDn=}\\
  &\mathcal{A}_{\mathcal{D}}(\bm{\lambda})^{\dagger}
  \phi_{\mathcal{D}\,n-1}(x;\bm{\lambda}+\bm{\delta})
  =\sqrt{B_{\mathcal{D}}(0;\bm{\lambda})}\,
  \phi_{\mathcal{D}\,n}(x;\bm{\lambda})
  \ \ (x=0,1,\ldots,x_{\text{max}}).
  \label{ADdphiDn=}
\end{align}
The forward and backward shift operators are defined by
\begin{align}
  \mathcal{F}_{\mathcal{D}}(\bm{\lambda})&\eqdef
  \sqrt{B_{\mathcal{D}}(0;\bm{\lambda})}\,
  \psi_{\mathcal{D}}\,(x;\bm{\lambda}+\bm{\delta})^{-1}\circ
  \mathcal{A}_{\mathcal{D}}(\bm{\lambda})\circ
  \psi_{\mathcal{D}}\,(x;\bm{\lambda})\n
  &=\frac{B(0;\bm{\lambda}+M\tilde{\bm{\delta}})}
  {\varphi(x;\bm{\lambda}+M\tilde{\bm{\delta}})
  \check{\Xi}_{\mathcal{D}}(x+1;\bm{\lambda})}
  \Bigl(\check{\Xi}_{\mathcal{D}}(x+1;\bm{\lambda}+\bm{\delta})
  -\check{\Xi}_{\mathcal{D}}(x;\bm{\lambda}+\bm{\delta})e^{\partial}\Bigr),
  \label{calFD}\\
  \mathcal{B}_{\mathcal{D}}(\bm{\lambda})&\eqdef
  \frac{1}{\sqrt{B_{\mathcal{D}}(0;\bm{\lambda})}}\,
  \psi_{\mathcal{D}}\,(x;\bm{\lambda})^{-1}\circ
  \mathcal{A}_{\mathcal{D}}(\bm{\lambda})^{\dagger}\circ
  \psi_{\mathcal{D}}\,(x;\bm{\lambda}+\bm{\delta})\n
  &=\frac{1}{B(0;\bm{\lambda}+M\tilde{\bm{\delta}})
  \check{\Xi}_{\mathcal{D}}(x;\bm{\lambda}+\bm{\delta})}
  \label{calBD}\\
  &\quad\times
  \Bigl(B(x;\bm{\lambda}+M\tilde{\bm{\delta}})
  \check{\Xi}_{\mathcal{D}}(x;\bm{\lambda})
  -D(x;\bm{\lambda}+M\tilde{\bm{\delta}})
  \check{\Xi}_{\mathcal{D}}(x+1;\bm{\lambda})e^{-\partial}\Bigr)
  \varphi(x;\bm{\lambda}+M\tilde{\bm{\delta}}),
  \nonumber
\end{align}
and their actions on $\check{P}_{\mathcal{D},n}(x;\bm{\lambda})$ are
\begin{equation}
  \mathcal{F}_{\mathcal{D}}(\bm{\lambda})
  \check{P}_{\mathcal{D},n}(x;\bm{\lambda})
  =\mathcal{E}_n(\bm{\lambda})
  \check{P}_{\mathcal{D},n-1}(x;\bm{\lambda}+\bm{\delta}),\quad
%  \label{FDPDn=}\\
  \mathcal{B}_{\mathcal{D}}(\bm{\lambda})
  \check{P}_{\mathcal{D},n-1}(x;\bm{\lambda}+\bm{\delta})
  =\check{P}_{\mathcal{D},n}(x;\bm{\lambda}).
  \label{BDPDn=}
\end{equation}
As in the original ($q$-)Racah theory \eqref{BPn-1=Pn}, these formulas
are useful for the explicit calculation of the multi-indexed polynomials.
The similarity transformed Hamiltonian is
\begin{align}
  \widetilde{\mathcal{H}}_{\mathcal{D}}(\bm{\lambda})
  &\eqdef\psi_{\mathcal{D}}(x;\bm{\lambda})^{-1}\circ
  \mathcal{H}_{\mathcal{D}}(\bm{\lambda})\circ
  \psi_{\mathcal{D}}(x;\bm{\lambda})
  =\mathcal{B}_{\mathcal{D}}(\bm{\lambda})
  \mathcal{F}_{\mathcal{D}}(\bm{\lambda})\n
  &=B(x;\bm{\lambda}+M\tilde{\bm{\delta}})\,
  \frac{\check{\Xi}_{\mathcal{D}}(x;\bm{\lambda})}
  {\check{\Xi}_{\mathcal{D}}(x+1;\bm{\lambda})}
  \biggl(\frac{\check{\Xi}_{\mathcal{D}}(x+1;\bm{\lambda}+\bm{\delta})}
  {\check{\Xi}_{\mathcal{D}}(x;\bm{\lambda}+\bm{\delta})}-e^{\partial}
  \biggr)\n
  &\quad+D(x;\bm{\lambda}+M\tilde{\bm{\delta}})\,
  \frac{\check{\Xi}_{\mathcal{D}}(x+1;\bm{\lambda})}
  {\check{\Xi}_{\mathcal{D}}(x;\bm{\lambda})}
  \biggl(\frac{\check{\Xi}_{\mathcal{D}}(x-1;\bm{\lambda}+\bm{\delta})}
  {\check{\Xi}_{\mathcal{D}}(x;\bm{\lambda}+\bm{\delta})}-e^{-\partial}
  \biggr),
\end{align}
and the multi-indexed orthogonal polynomials
$\check{P}_{\mathcal{D},n}(x;\bm{\lambda})$ are its eigenpolynomials:
\begin{equation}
  \widetilde{\mathcal{H}}_{\mathcal{D}}(\bm{\lambda})
  \check{P}_{\mathcal{D},n}(x;\bm{\lambda})=\mathcal{E}_n(\bm{\lambda})
  \check{P}_{\mathcal{D},n}(x;\bm{\lambda}).
  \label{tHPDn=}
\end{equation}
Other intertwining relations are
\begin{align*}
  \kappa^{\frac12}
  \hat{\mathcal{A}}_{d_1\ldots d_{s+1}}(\bm{\lambda}+\bm{\delta})
  \mathcal{A}_{d_1\ldots d_s}(\bm{\lambda})
  &=\mathcal{A}_{d_1\ldots d_{s+1}}(\bm{\lambda})
  \hat{\mathcal{A}}_{d_1\ldots d_{s+1}}(\bm{\lambda}),
%  \label{hAA=AhA}\\
  \\
  \kappa^{-\frac12}
  \hat{\mathcal{A}}_{d_1\ldots d_{s+1}}(\bm{\lambda})
  \mathcal{A}_{d_1\ldots d_s}(\bm{\lambda})^{\dagger}
  &=\mathcal{A}_{d_1\ldots d_{s+1}}(\bm{\lambda})^{\dagger}
  \hat{\mathcal{A}}_{d_1\ldots d_{s+1}}(\bm{\lambda}+\bm{\delta}),
%  \label{hAAd=AdhA}
\end{align*}
with the potential functions given in \eqref{Bdsform}--\eqref{Ddsform}
(with $s\to s+1$).

Including the level 0 deletion corresponds to $M-1$ virtual
states deletion:
\begin{equation}
  \check{P}_{\mathcal{D},n}(x;\bm{\lambda})\Bigm|_{d_M=0}
  =\check{P}_{\mathcal{D}',n}(x;\bm{\lambda}+\tilde{\bm{\delta}}),\quad
  \mathcal{D}'=\{d_1-1,\ldots,d_{M-1}-1\}.
  \label{dM=0}
\end{equation}
This formula is similar to those in the multi-indexed Jacobi theory,
eqs.(48)--(49) in \cite{os25}.
The denominator polynomial $\Xi_{\mathcal{D}}$ behaves similarly.
This is why we have restricted $d_j\geq 1$.

The coefficients of the highest degree terms of the polynomials
$\Xi_{\mathcal{D}}$ and $P_{\mathcal{D},n}$,
\begin{align*}
  \Xi_{\mathcal{D}}(y;\bm{\lambda})&=c^{\Xi}_{\mathcal{D}}(\bm{\lambda})
  y^{\ell}+(\text{lower order terms}),\\
  P_{\mathcal{D},n}(y;\bm{\lambda})&=c^{P}_{\mathcal{D},n}(\bm{\lambda})
  y^{\ell+n}+(\text{lower order terms}),
\end{align*}
are
\begin{align}
  &c^{\Xi}_{\mathcal{D}}(\bm{\lambda})=\left\{
  \begin{array}{ll}
  {\displaystyle
  \frac{\prod_{j=1}^M(-a-b+c+d+d_j+1)_{d_j}}
  {\prod_{1\leq j<k\leq M}(-a-b+c+d+d_j+d_k+1)}
  \prod_{j=1}^M\frac{(c,d-a+1,d-b+1)_{j-1}}{(c,d-a+1,d-b+1)_{d_j}}
  }&:\text{R}\\
  {\displaystyle
  \frac{\prod_{j=1}^M(a^{-1}b^{-1}cdq^{d_j+1};q)_{d_j}}
  {\prod_{1\leq j<k\leq M}(1-a^{-1}b^{-1}cdq^{d_j+d_k+1})}
  \prod_{j=1}^M\frac{(c,a^{-1}dq,b^{-1}dq;q)_{j-1}}
  {(c,a^{-1}dq,b^{-1}dq;q)_{d_j}}
  }&:\text{$q$R}
  \end{array}\right.,\n
%  \label{c_XiD}\\
%
  &c^P_{\mathcal{D},n}(\bm{\lambda})=c^{\Xi}_{\mathcal{D}}(\bm{\lambda})
  \times\left\{
  \begin{array}{ll}
  {\displaystyle
  \frac{(a+b+c-d+n-1)_n\,(c)_M}
  {(a,b,c)_n\prod_{j=1}^M(c+n+d_j)}
  }&:\text{R}\\[12pt]
  {\displaystyle
  \frac{(abcd^{-1}q^{n-1};q)_n\,(c;q)_M}
  {(a,b,c;q)_n\prod_{j=1}^M(1-cq^{n+d_j})}
  }&:\text{$q$R}
  \end{array}\right..
  \label{c_PDn}
\end{align}

The exceptional $X_{\ell}$ ($q$-)Racah orthogonal polynomials presented in
\cite{os23} correspond to the simplest case $M=1$, $\mathcal{D}=\{\ell\}$,
$\ell\geq 1$:
\begin{equation}
  \check{\xi}_{\ell}(x;\bm{\lambda})
  =\check{\Xi}_{\{\ell\}}
  (x;\bm{\lambda}+\ell\bm{\delta}-\tilde{\bm{\delta}}),\quad
  \check{P}_{\ell,n}(x;\bm{\lambda})
  =\check{P}_{\{\ell\},n}(x;\bm{\lambda}+\ell\bm{\delta}-\tilde{\bm{\delta}}).
\end{equation}

%%%%%%%%%%%%%%%%%%%%%%%%%%%%%%%%%%%%%%%%%%%%%%%%%%%%%%%%%%%%%%%
%                                                             %
%  4. Summary and Comments                                    %
%                                                             %
%%%%%%%%%%%%%%%%%%%%%%%%%%%%%%%%%%%%%%%%%%%%%%%%%%%%%%%%%%%%%%%
\section{Summary and Comments}
\label{summary}
\setcounter{equation}{0}

Following the examples of multi-indexed Laguerre and Jacobi polynomials
\cite{os25}, multi-indexed ($q$-)Racah polynomials, the discrete
quantum mechanics counterparts, are constructed.
These new polynomials could be considered as a further generalisation
of Bannai-Ito polynomials \cite{bannai}.
The next stage will be the construction of multi-indexed Askey-Wilson
and Wilson polynomials. The basic logic is the same for the ordinary
quantum mechanics as well as for the discrete quantum mechanics with
real \cite{os12} or pure imaginary shifts \cite{os13}.
Starting from the factorised Hamiltonians of exactly solvable quantum
mechanical systems, a series of new `deformed' exactly solvable quantum
systems are generated by applying Crum-Krein-Adler formulas
\cite{crum,adler} or multiple Darboux transformations \cite{darb}
through deletion of various virtual states instead of eigenstates.
The virtual state vectors are polynomial `solutions' of a virtual
Hamiltonian which is obtained by twisting the discrete symmetry of
the original Hamiltonian. They fail to satisfy the Schr\"odinger equation
of the virtual Hamiltonian at one of the boundaries, at $x=x_{\text{max}}$.
When there is only one extra index $\mathcal{D}=\{\ell\}$ ($\ell\ge1$),
the multi-indexed ($q$-)Racah polynomials reduce to the exceptional
polynomials \cite{os23,os24}.
Like the exceptional polynomials, the multi-indexed ($q$-)Racah polynomials
do not satisfy the three term recurrence relations. On the other hand,
their dual polynomials satisfy the three term recurrence relations
because of the tri-diagonal form of the Hamiltonian.
As for the parameter ranges in which \eqref{B'>0,..}--\eqref{D'>0,..},
\eqref{xi>0}--\eqref{tEv<0} are satisfied, we have taken
conservative ones, \eqref{pararange}--\eqref{pararangeq}, \eqref{Mrange},
\eqref{vrange}.
It is quite possible that the valid parameter ranges could be enlarged.
The difference equations for the multi-indexed ($q$-)Racah polynomials,
\eqref{BDPDn=}, \eqref{tHPDn=} are purely algebraic and they hold for
any parameter ranges.

As in the ordinary Sturm-Liouville case (the oscillation theorem)
the multi-indexed orthogonal polynomial $P_{\mathcal{D},n}(y;\bm{\lambda})$
has $n$ zeros in the orthogonality range,
$0<y<\eta(x_{\text{max}};\bm{\lambda}+M\tilde{\bm{\delta}})$.
This is a general property of the eigenvectors of a Jacobi matrix
of the form \eqref{Hdef}, \eqref{Hprime} etc. See \cite{gladwell}. 

A few words on the mirror reflection with respect to the mid point
of the $x$-grid, $x\to x_\text{max}-x=N-x$. The original ($q$-)Racah
polynomial under the reflection is described by the same polynomial
with different parameters:
\begin{align}
  \check{P}_n(N-x;\bm{\lambda})
  &=A\times\check{P}_n\bigl(x;(\lambda_1,\lambda_1+\lambda_3-\lambda_4,
  \lambda_1+\lambda_2-\lambda_4,2\lambda_1-\lambda_4)\bigr),\n
  A&=\check{P}_n(N;\bm{\lambda})
  =\left\{
  \begin{array}{ll}
  \displaystyle{\frac{(a+b-d,a+c-d)_n}{(b,c)_n}}&:\text{R}\\[10pt]
  \displaystyle{\bigl(a^{-1}d\bigr)^n\,
  \frac{(abd^{-1},acd^{-1};q)_n}{(b,c;q)_n}}&:\text{$q$R}
  \end{array}\right..
\end{align}
This corresponds to the mirror reflection formula of the Jacobi polynomials:
\begin{equation*}
  P_n^{(\alpha,\beta)}(-\eta)=(-1)^nP_n^{(\beta,\alpha)}(\eta),\quad
  \eta(x)=\cos 2x,\quad \eta(\tfrac{\pi}{2}-x)=-\eta(x).
\end{equation*}
The type $\I$ and $\II$ virtual state wavefunctions for the Jacobi case
have a twisted boundary condition at $x=\frac{\pi}{2}$ and $x=0$, respectively
\cite{os25} and their polynomial parts are related by this mirror reflection
\cite{os19}.
The virtual state vectors, which fail to satisfy the equation at
$x=x_\text{max}$, correspond to the type $\I$.
By the mirror reflection, one can consider the type $\II$ virtual state vectors,
which fail to satisfy the equation at $x=0$, instead of $x=x_\text{max}$.
By using these type $\II$ virtual state vectors, the mirror reflexed version of
the multi-indexed ($q$-)Racah polynomials can be constructed.
They are related to the multi-indexed ($q$-)Racah polynomials by
\begin{align}
  \check{P}_{\mathcal{D},n}^{\text{mirror}}(x;\bm{\lambda})
  &=A\times \check{P}_{\mathcal{D},n}\bigl(N-x;
  (\lambda_1,\lambda_1+\lambda_3-\lambda_4,\lambda_1+\lambda_2-\lambda_4,
  2\lambda_1-\lambda_4)\bigr),\n
  A^{-1}&=\check{P}_{\mathcal{D},n}\bigl(N;
  (\lambda_1,\lambda_1+\lambda_3-\lambda_4,\lambda_1+\lambda_2-\lambda_4,
  2\lambda_1-\lambda_4)\bigr).
\end{align}
The normalisation conditions are $\check{P}_{n}(0;\bm{\lambda})=
\check{P}_{\mathcal{D},n}(0;\bm{\lambda})
=\check{P}_{\mathcal{D},n}^{\text{mirror}}(0;\bm{\lambda})=1$.
Contrary to the Jacobi case \cite{os25}, the type $\I$ and $\II$ virtual
state vectors cannot be used together to generate new multi-indexed
($q$-)Racah polynomials.

Various orthogonal polynomials are obtained from the ($q$-)Racah
polynomials in certain limits.
Similarly, from the multi-indexed ($q$-)Racah polynomials presented
in the previous section, we can obtain the multi-indexed version of
various orthogonal polynomials, such as the ($q$-)Hahn,
dual ($q$-)Hahn, alternative $q$-Hahn (see \S\,V.C.1 of \cite{os12}), etc.
The infinite dimensional cases, the little $q$-Jacobi, ($q$-)Meixner,
etc will be reported in a separate publication.
In certain limiting processes, the mirror reflection does not commute with
the limit, as is well known in the Jacobi $\to$ Laguerre limits
\cite{os19,os25}. The mirror reflexed multi-indexed polynomials are supposed
to play certain roles in such limits.

Let us emphasise that the discrete symmetries of the original
($q$-)Racah systems and their twisting, which are essential
for the construction of virtual Hamiltonians and virtual state
vectors, are easily recognised in the present parametrisation
\eqref{Bform}--\eqref{Dform} \cite{os12}, but rather unclear
in the original parametrisation \cite{askey,ismail,koeswart}.
This is a good reason to promote the ($q$-)Racah systems in our
parametrisation.

%%%%%%%%%%%%%%%%%%%%%%%%%%%%%%%%%%%%%%%%%%%%%%%%%%%%%%%%%%%%%%%
%                                                             %
%  Acknowledgments                                            %
%                                                             %
%%%%%%%%%%%%%%%%%%%%%%%%%%%%%%%%%%%%%%%%%%%%%%%%%%%%%%%%%%%%%%%
\section*{Acknowledgements}

We thank a referee for many useful comments.
S.\,O. thanks to his late father Zen Odake for warm encouragement.
R.\,S. is supported in part by Grant-in-Aid for Scientific Research
from the Ministry of Education, Culture, Sports, Science and Technology
(MEXT), No.23540303 and No.22540186.

%%%%%%%%%%%%%%%%%%%%%%%%%%%%%%%%%%%%%%%%%%%%%%%%%%%%%%%%%%%%%%%
%                                                             %
%  References                                                 %
%                                                             %
%%%%%%%%%%%%%%%%%%%%%%%%%%%%%%%%%%%%%%%%%%%%%%%%%%%%%%%%%%%%%%%

\end{document}